\documentclass[review, 12pt]{elsarticle}
\usepackage[english]{babel}
\usepackage[T1]{fontenc}
\usepackage[latin9]{inputenc}
\usepackage{geometry}
\usepackage{float}
\usepackage{amscd}
\usepackage{amsmath}
\usepackage{amssymb}
\usepackage{stmaryrd}
\usepackage{graphicx}
\usepackage{setspace}
\usepackage{framed}
\usepackage{algpseudocode}
\usepackage{caption}
\DeclareCaptionType{algorithm}

\makeatletter
\def\ps@pprintTitle{%
   \let\@oddhead\@empty
   \let\@evenhead\@empty
   \def\@oddfoot{\reset@font\hfil\thepage\hfil}
   \let\@evenfoot\@oddfoot
}
\makeatother

\makeatletter

\floatstyle{ruled}
\usepackage[symbol]{footmisc}
\date{}

\@ifundefined{showcaptionsetup}{}{%
 \PassOptionsToPackage{caption=false}{subfig}}
\usepackage{subfig}
\makeatother

\usepackage{babel}
\begin{document}

\begin{frontmatter}

\title{3D Reconstruction of the Magnetic Vector Potential using
Model Based Iterative Reconstruction}

\author[cmu]{Prabhat KC}
\author[pur]{K. Aditya Mohan}
\author[anl]{Charudatta Phatak}
\author[pur]{Charles Bouman}
\author[cmu]{Marc De Graef\corref{cor1}}
\ead{degraef@cmu.edu }
\address[cmu]{Department of Materials Science and Engineering, Carnegie Mellon University, Pittsburgh, Pennsylvania, USA}
\address[pur]{Department of Electrical and Computer Engineering, Purdue University, West Lafayette, IN, USA}
\address[anl]{Argonne National Laboratory, Lemont, Illinois, USA}
\cortext[cor1]{Corresponding author. Tel: +1 412 268 8527; fax: +1 412 268 759}

\begin{abstract}
Lorentz Transmission Electron Microscopy (TEM) observations
of magnetic nanoparticles contain information on the magnetic and electrostatic
potentials. Vector Field Electron Tomography
(VFET) can be used to reconstruct electromagnetic potentials
of the nanoparticles from their corresponding LTEM images. 
The VFET approach is based on the conventional filtered back projection 
approach to tomographic reconstructions and the availability
of an incomplete set of measurements due to experimental limitations
means that the reconstructed vector fields exhibit significant
artifacts. In this paper, we outline a model-based iterative
reconstruction (MBIR) algorithm to reconstruct the magnetic vector
potential of magnetic nanoparticles. We combine a forward model for image
formation in TEM experiments with a prior model to formulate the tomographic
problem as a maximum a-posteriori probability estimation problem (MAP).
The MAP cost function is minimized iteratively to determine the vector
potential. A comparative reconstruction study of simulated as well
as experimental data sets show that the MBIR approach yields quantifiably 
better reconstructions than the VFET approach. 
\end{abstract}


\end{frontmatter}

%%%%%%%%%%%%%%%%%%%%%%%%%%%%%
\section{Introduction\label{sec:Introduction}}
%%%%%%%%%%%%%%%%%%%%%%%%%%%%%
In a Lorentz Transmission Electron Microscopy (LTEM) experiment,
an electron propagating through a thin specimen experiences a Lorentz
Force $\mathbf{F}_{L}=-e(\mathbf{E}+\mathbf{v}\times\mathbf{B})$
\cite{lorentz_force_jackson} due to the sample's electrostatic
field, $\mathbf{E}$, and magnetic field, $\mathbf{B}$; $-e$ is
the electron's charge and $\mathbf{v}$ its velocity. This (classical)
force generates a deflection of the electron trajectory, which can
be used to explain the Fresnel and Foucault observation modes \cite{marc_editor_book}. A more
robust explanation of the nature of the electron-specimen interaction
involves quantum mechanics, in which the electron is described by
a wave function $\psi(\mathbf{r}_{\perp})=a(\mathbf{r}_{\perp})e^{i\varphi(\mathbf{r_{\perp}})}$
\cite{marc_editor_book}. Elastic scattering in the sample produces
variations of the amplitude $a(\mathbf{r}_{\perp})$, whereas the
electromagnetic potentials affect the phase $\varphi(\mathbf{r}_{\perp})$
of the wave; $\mathbf{r}_{\perp}$ is a vector normal to the propagation
direction. Aharonov and Bohm \cite{A_B_paper} showed, in 1959, that
the phase of the exit wave function encodes information on the sample\textquoteright s
electrostatic potential, $V(\mathbf{r}_{\perp},z)$, and magnetic
vector potential, $\mathbf{A}(\mathbf{r}_{\perp},z)$, as follows:
\begin{equation}
\varphi(\mathbf{r}_{\perp})=\varphi_{e}(\mathbf{r}_{\perp})+\varphi_{m}(\mathbf{r}_{\perp})=\frac{\pi}{\lambda E_{t}}\underset{L}{\int}V(\mathbf{r}_{\perp},z)dz-\frac{e}{\hslash}\underset{L}{\int}\mathbf{A}(\mathbf{r}_{\perp},z)\cdot d\mathbf{r},\label{eq:A_B phase shift eq}
\end{equation}
where $\hslash$ is the reduced Planck's constant, $E_{t}$ is
the total beam energy, and the
integrals are carried out along the beam direction, $L$. The total phase
shift, $\varphi$, consists of an electrostatic contribution, $\varphi_{e}$,
and a magnetic contribution, $\varphi_{m}$. The phases are not directly
observable, but their effect on the image contrast can be determined
by considering the point spread function, $\mathcal{T}_{L}(\mathbf{r}_{\perp})$,
of the Lorentz lens. The image intensity is then given by the modulus-squared
of the convolution product $\psi(\mathbf{r}_{\perp})\varotimes\mathcal{T}_{L}(\mathbf{r_{\perp}})$
\cite{marc_editor_book}. Hence, characterization of the electromagnetic
potentials begins with phase shift retrieval from the
image intensities, using either electron holography \cite{tomography_holography_review}
or the transport-of-intensity equation (TIE) formalism \cite{TIE_paper},
which is based on a through-focus series of Fresnel images. We use
the linearity of the TIE \cite{emma_TIE_paper} and time reversal symmetry
to retrieve the individual phases $\varphi_{e}$ and $\varphi_{m}$. 

\begin{figure}[h]
\centering\leavevmode
\includegraphics[width=0.5\textwidth]{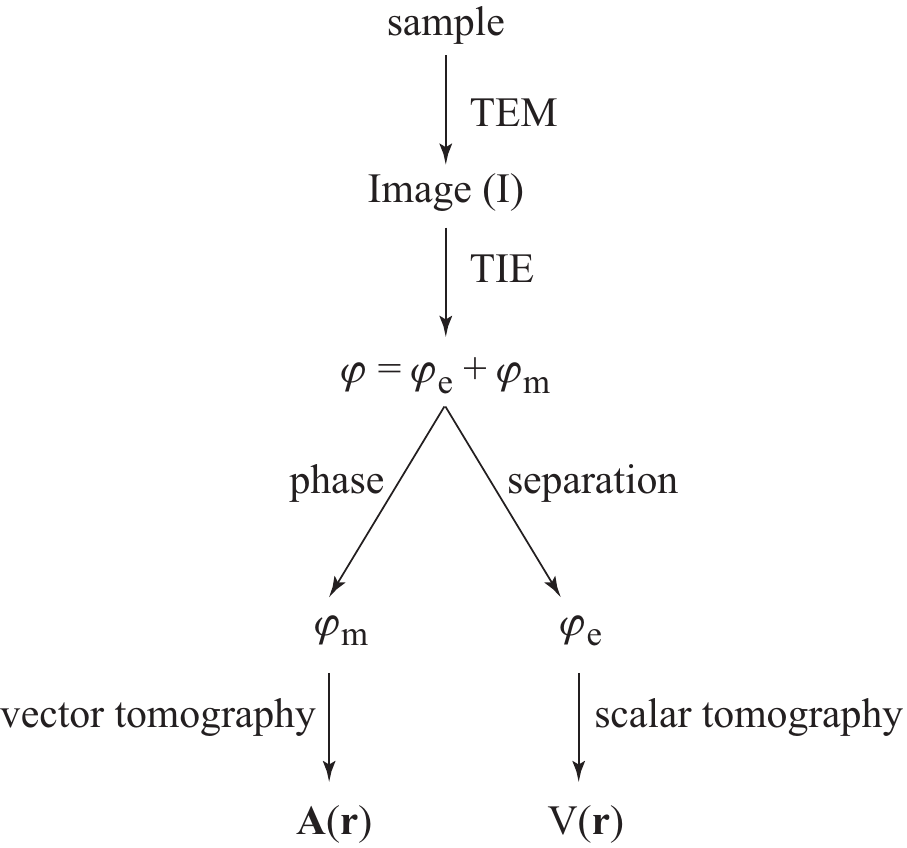}
\caption{A flow chart illustrating the methodology to determine electromagnetic
potentials of a magnetic nanoparticle sample. \label{fig:TEM_electromag_flowchart}}
\end{figure}

Characterization of the electromagnetic fields is then achieved 
by performing scalar field and vector field tomographic reconstructions to determine
$V(\mathbf{r})$ and $\mathbf{A}(\mathbf{r})$, respectively. We refer
to the use of vector field tomography to reconstruct the electromagnetic potentials
as Vector Field Electron Tomography (VFET) \cite{vfet_noise}. A schematic of the
operations performed to complete an electromagnetic characterization
task is shown in Fig.~\ref{fig:TEM_electromag_flowchart}.

In this contribution, we primarily focus on vector field
tomography to reconstruct $\mathbf{A}(\mathbf{r})$. In recent years, the VFET 
approach employed the filtered back projection (FBP) approach to perform the reconstructions \cite{vfet_doppler,vfet_three_tilt}. Although FBP yields a good estimate with a complete set
of measurements, the typical missing wedge of TEM data significantly diminishes 
the quality of the reconstructions \cite{e_tomo_review}.
In addition, typical tilt series are obtained using an angular step size of
$2^{\circ}$--$5^{\circ}$ to minimize the necessary pre-processing steps (image alignments)
and reduce beam damage. These limitations, collectively,
yield a reconstruction result that can exhibit substantial artifacts. 
To alleviate these problems, we resort to a more robust and statistically
based tomographic reconstruction framework known as Model-Based Iterative
Reconstruction (MBIR) to determine $\mathbf{A}(\mathbf{r})$. This approach has 
had considerable success in improving reconstruction quality in scalar tomography
\cite{c_bouman_local_strategy,qGGMRF_proof}. 

In section~\ref{sec:VFET_framework}, we briefly outline conventional VFET and show
how we can reconstruct all three component of $\mathbf{A}(\mathbf{r})$
from just two tilt series; we also perform an error analysis of the
quality of VFET reconstructions in the presence of a missing wedge. Next, 
in section~\ref{sec:gen-MBIR-framework}, we provide an overview of
the MBIR framework, and in section~\ref{sec:MBIR_vec_imp}, we incorporate
MBIR into the reconstruction of $\mathbf{A}(\mathbf{r})$ and compare
the results with those from conventional VFET reconstructions.

%%%%%%%%%%%%%%%%%%%%%%%%%%%%%
\section{Vector Field Electron Tomography\label{sec:VFET_framework}}
%%%%%%%%%%%%%%%%%%%%%%%%%%%%%
Vector field tomography is relatively new; it was not until 1988 when Norton \cite{norton_paper}, for the first
time, outlined a mathematical model to determine the 2D fluid field
from acoustic time travel measurements. Subsequent years saw extensions
of 2D vector tomography to 3D cases; in particular, Juhlin \cite{Juhin_paper}
resolved the solenoidal part of a divergence free flow field using
ultrasound Doppler measurements in 1992. In 2005, Lade
et al.\,\cite{vfet_doppler} presented the VFET model to reconstruct
3D vector fields from longitudinal and transverse measurements.
In 2008, Phatak et al.\ \cite{vfet_cd} used the VFET approach
to reconstruct the magnetic vector potential and induction
of magnetic nanoparticles. Since this VFET model is still relevant for our
new MBIR method, we devote this section to a brief review of the VFET framework.

\begin{figure}[ht]
\centering\leavevmode
\includegraphics[width=0.4\textwidth]{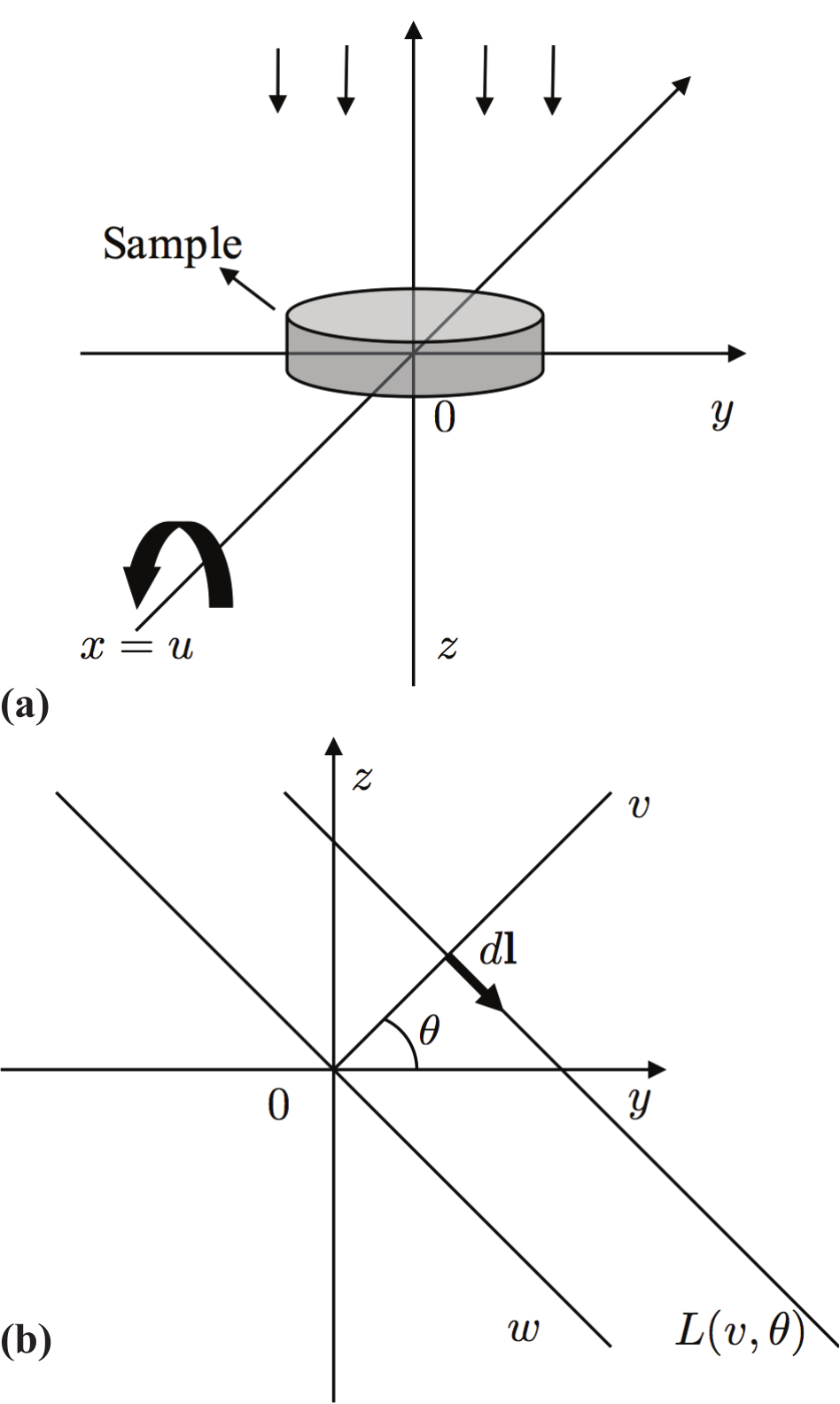}
\caption{(a) Illustration of phase shift acquisition for an $x$-tilt series with
the arrows representing the electron propagation direction
and the curved arrow indicating counter-clockwise sample rotation.
(b) Representation of the reference frame used to express the differential vector element $\mathrm{d}\mathbf{l}$.\label{fig:vector_element_fig}}
\end{figure}

Since tomographic reconstructions require a forward model to project the object being reconstructed,
we begin by considering the computation of the magnetic phase shift.
The relation $\varphi_{m}(\mathbf{r}_{\perp})=-\frac{e}{\hslash}\underset{L}{\int}\mathbf{A}(\mathbf{r}_{\perp},z)\cdot \mathrm{d}\mathbf{r}$
describes the magnetic phase shift obtained at $0^{\circ}$ tilt.
To obtain the phase shift for a tilted sample we consider a tilt series around the $x$ axis (counterclockwise);
the new coordinate vectors, $\mathbf{t},$
can be expressed in terms of the original ones, $\mathbf{r},$ by $\mathbf{r}=R_{\theta,x}\mathbf{t}$
where $\mathbf{r}=[x\ y\ z]$, $\mathbf{t}=[u\ v\ w]$,
and $R_{\theta,x}$ is the counter-clockwise rotation matrix (Fig.\,\ref{fig:vector_element_fig}(a)).
From Fig.\,\ref{fig:vector_element_fig}(b), the vectorial line element,
$\mathrm{d}\mathbf{l}$, of the projection line $L(v,\theta)$, can be written
as $\mathrm{d}\mathbf{l}=[\hat{y}\sin(\theta)-\hat{z}\cos(\theta)]\mathrm{d}l$.

Writing $\mathbf{A}(\mathbf{r})=A_{x}(x,y,z)\hat{x}+A_{y}(x,y,z)\hat{y}+A_{z}(x,y,z)\hat{z}$,
a generic projection equation for the $x$ tilt series in Fourier
space, $\tilde{\varphi}_{m,x}$, can be obtained as:
\begin{equation}
\tilde{\varphi}_{m,x}(k_{u},k_{v})=-\sin\theta\tilde{A}_{y}(k_{u},k_{v}\cos\theta,k_{v}\sin\theta)+\cos\theta\tilde{A}_{z}(k_{u},k_{v}\cos\theta,k_{v}\sin\theta);\label{eq:x_tilt_fst}
\end{equation}
a similar analysis for the $y$ tilt series produces
\begin{equation}
\tilde{\varphi}_{m,y}(k_{u},k_{v})=-\sin\theta\tilde{A}_{x}(k_{u}\cos\theta,k_{v},k_{u}\sin\theta)+\cos\theta\tilde{A}_{z}(k_{u}\cos\theta,k_{v},k_{u}\sin\theta).\label{eq:y_tilt_fst}
\end{equation}
Eqs.\,\ref{eq:x_tilt_fst} and \ref{eq:y_tilt_fst}
represent the Fourier slice theorem for 3D vector fields for $x$
tilt series and $y$ tilt series respectively. 

The formulation of the reconstruction procedure
by means of the VFET approach begins by imposing a gauge constraint
on the magnetic vector potential, i.e., $\nabla\cdot\mathbf{A}=0$. This
constraint is written in Fourier space as:
\begin{equation}
\mathbf{k}.\tilde{\mathbf{A}}=k_{x}\tilde{A}_{x}+k_{y}\tilde{A}_{y}+k_{z}\tilde{A}_{z}=0.\label{eq:coulomb_gauze}
\end{equation}
Combining eqs.\,\ref{eq:x_tilt_fst}, \ref{eq:y_tilt_fst},
and \ref{eq:coulomb_gauze} and solving for the components of $\mathbf{A}(\mathbf{k})$
one obtains:
\begin{align}
\tilde{A}_{x}&=\frac{k_{y}k_{x}k_{v}\tilde{\varphi}_{m,x}-k_{u}(k_{y}^{2}+k_{z}^{2})\tilde{\varphi}_{m,y}}{k_{z}(k_{x}^{2}+k_{y}^{2}+k_{z}^{2})};\label{eq:ax_fs_dm}\\
\tilde{A}_{y}&=\frac{-(k_{x}^{2}+k_{z}^{2})k_{v}\tilde{\varphi}_{m,x}+k_{v}k_{x}k_{y}\tilde{\varphi}_{m,y}}{k_{z}(k_{x}^{2}+k_{y}^{2}+k_{w}^{2})};\label{eq:ay_fs_dm}\\
\tilde{A}_{z}&=\frac{k_{y}k_{z}k_{v}\tilde{\varphi}_{m,x}-k_{u}k_{x}k_{z}\tilde{\varphi}_{m,y}}{k_{z}(k_{x}^{2}+k_{y}^{2}+k_{w}^{2})}.\label{eq:az_fs_dm}
\end{align}
In principle, one could interpolate the projection measurements from the polar grid to a Cartesian grid and perform an inverse
Fourier transform (FT) to solve for $\mathbf{A}(\mathbf{r})$. However, this is a numerically unstable approach and the filtered back projection formulae, as outlined in ref.\ \cite{kak_slaney}, facilitate a more stable reconstruction. In the present case, the FBP formulae become (the asterisk subscript represents $x$, $y$, or $z$):
\begin{equation}
A_{*}(x,y,z)=\underset{0}{\overset{\pi}{\int}}\left[\varphi_{m,x,*}(x,y\cos\theta+z\sin\theta)+\varphi_{m,y,*}(x\cos\theta+z\sin\theta,y)\right]d\theta,\label{eq:Ar_fbp_formula}
\end{equation}
where
\begin{align}
\varphi_{m,x,*}&=\mathcal{F}_{2}^{-1}\left[\frac{|k_{v}|}{\sin\theta(k_{u}^{2}+k_{v}^{2})}\left[\begin{array}{c}
k_{u}k_{v}\cos\theta\\
-(k_{u}^{2}+k_{v}^{2}\sin\theta)\\
k_{v}^{2}\cos\theta\sin\theta
\end{array}\right]\tilde{\varphi}_{m,x}(k_{u},k_{v})\right]=\left[\begin{array}{c}
\varphi_{m,x,x}\\
\varphi_{m,x,y}\\
\varphi_{m,x,z}
\end{array}\right],\label{eq:phi_x_contribution_Ar}\\
\varphi_{m,y,*}&=\mathcal{F}_{2}^{-1}\left[\frac{|k_{u}|}{\sin\theta(k_{u}^{2}+k_{v}^{2})}\left[\begin{array}{c}
-(k_{v}^{2}+k_{u}^{2}\sin\theta)\\
k_{u}k_{v}\cos\theta\\
k_{u}^{2}\cos\theta\sin\theta
\end{array}\right]\tilde{\varphi}_{m,y}(k_{u},k_{v})\right]=\left[\begin{array}{c}
\varphi_{m,y,x}\\
\varphi_{m,y,y}\\
\varphi_{m,y,z}
\end{array}\right].\label{eq:phi_y_contribution_Ar}
\end{align}
In eqs.~\ref{eq:phi_x_contribution_Ar} and \ref{eq:phi_y_contribution_Ar},
$\vert k_u\vert$ and $\vert k_v\vert$ represent the filtering operation and $\mathcal{F}_{2}^{-1}$ 
the 2D inverse FT. The real-space vector quantities $\varphi_{m,x,*}$ and $\varphi_{m,y,*}$, are then used to
determine the three components of $\mathbf{A}(\mathbf{r})$ through an inverse Hough transform. 
The gauge condition on $\mathbf{A}$ will be incorporated into the MBIR approach as well.
However, we will replace the FBP formula of eq.~\ref{eq:Ar_fbp_formula}
with a Bayesian inference-based objective function which will be described in more detail in the next section. 

\begin{figure}[t]
\centering\leavevmode
\includegraphics[width=0.5\textwidth]{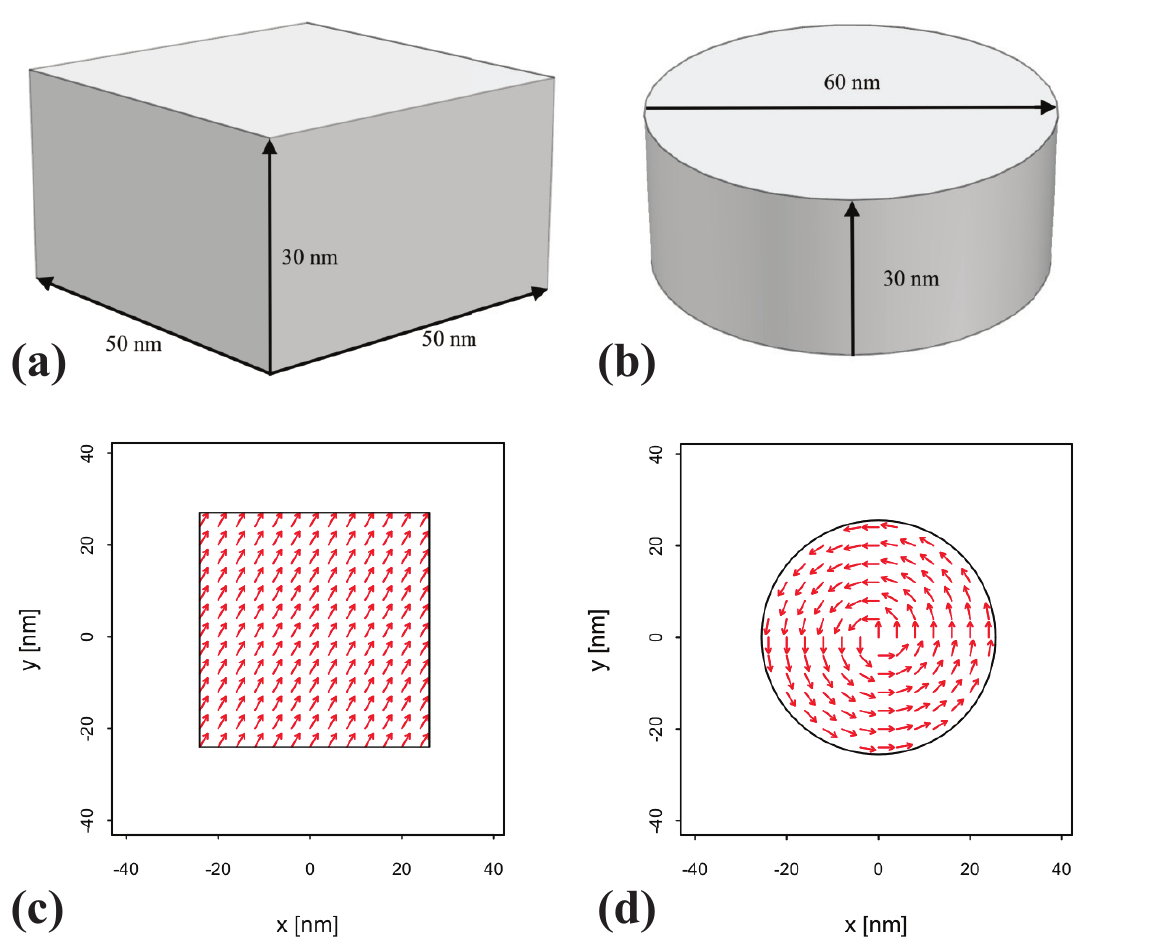}
\caption{(a) Prismatic MNP; (b) cylindrical MNP; (c) Vector plot showing uniform magnetization
(identical in each $z$-plane for a magnetization
direction of $\hat{\mathbf{m}}=[\cos\frac{\pi}{6},\sin\frac{\pi}{6},0]$.
(d) Vector plot illustrating a counter-clockwise vortex state in each $z$-plane of the cylindrical MNP. \label{fig:NP_mag}}
\end{figure}

Before we outline the MBIR framework, we illustrate the numerical implementation of
the VFET approach using uniformly and non-uniformly magnetized
simulated magnetic nanoparticles (MNPs); in particular, we use a uniformly magnetized prismatic particle
and a circular disk with a vortex state. The prismatic nanoparticle (NP) has dimensions of $\left[50\times50\times30\right]$ nm
and magnetization direction, $\mathbf{m}=\left[\cos\frac{\pi}{6},\sin\frac{\pi}{6},0\right]$. The circular disk has a diameter
of $60$ nm, a height of $30$ nm, and exhibits a counterclock-wise magnetization state with a sharp vortex \cite{philMagI_NP,emma_spherical_proj}. Both particles have a saturation magnetization of $B_{0}=1$ T. 
The two geometries and their magnetization states are depicted in Fig.~\ref{fig:NP_mag}. 

Starting from the magnetization states, the Fourier transform of the magnetic vector potential
can be expressed in terms of the shape function formalism of Ref.\,\cite{philMagI_NP} as:
\begin{equation}
\mathbf{A}(\mathbf{k})=-\frac{\mathrm{i}B_{0}}{k^{2}}D(\mathbf{k})(\hat{\mathbf{m}}\times\mathbf{k}),\label{eq:Ar_formula}
\end{equation}
where $D(\mathbf{k})$ is the shape amplitude. This expression allows for the computation of the 
magnetic phase shifts for two orthogonal tilt series around the $x$ and $y$ axes.
In one series we use projections over the full tilt range $[-90^{\circ},+90^{\circ}]$
with a $2^{\circ}$ step size while in a second series, with a missing wedge,
we use the sub-range of $[-70^{\circ},+70^{\circ}]$. After application of the VFET approach described above
we compare the reconstructed potentials with their corresponding known values (i.e., ground truths),
in terms of a normalized root mean square error (NRMSE)
between the reconstructed result, $\hat{\phi}$, and the ground truth, $\phi$,
as ($\phi$ represents one of the components of the vector potential):
\begin{equation}
    \text{NRMSE}=\frac{1}{\phi_{\text{max}}-\phi_{min}}\sqrt{\frac{1}{N}\underset{i=1}{\overset{N}{\sum}}(\hat{\phi_{i}}-\phi_{i})^{2}}.\label{eq:NRMSE}
\end{equation}

\begin{figure}[t]
\centering\leavevmode
\includegraphics[width=0.5\textwidth]{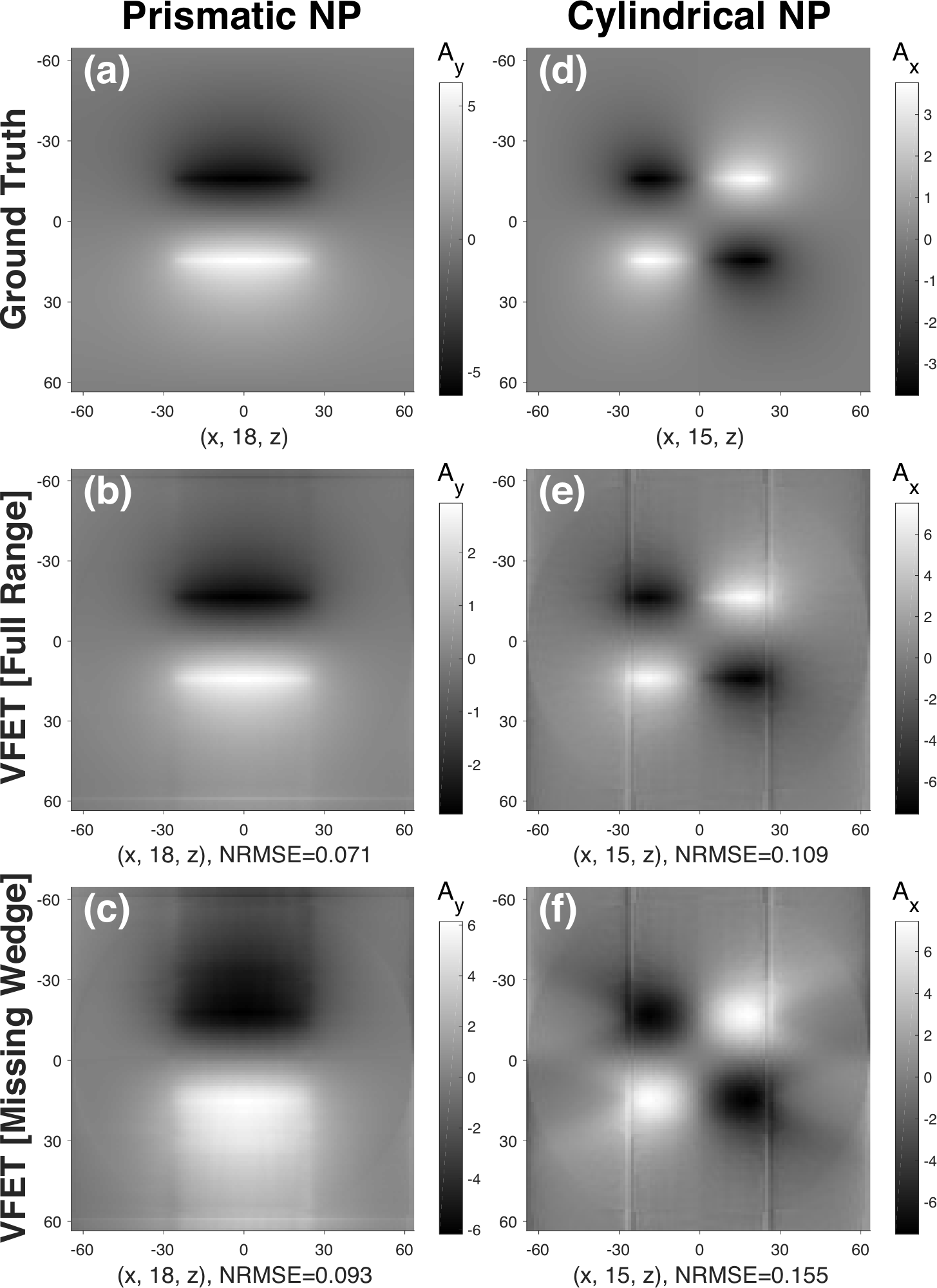}
\caption{Reconstructed magnetic vector potential from full range [center
row] and missing wedge [bottom row] projection data
sets. The top row depicts the ground truth or theoretical vector
components. The left column plots correspond to the plane $(x,18,z)$ of the prismatic
NP while the right column plots correspond to the plane $(x,15,z)$ of the cylindrical
NP. \label{fig:vfet_70vs90p_qualitative}}
\end{figure}

\begin{figure}[t]
\leavevmode\centering
\includegraphics[width=0.5\textwidth]{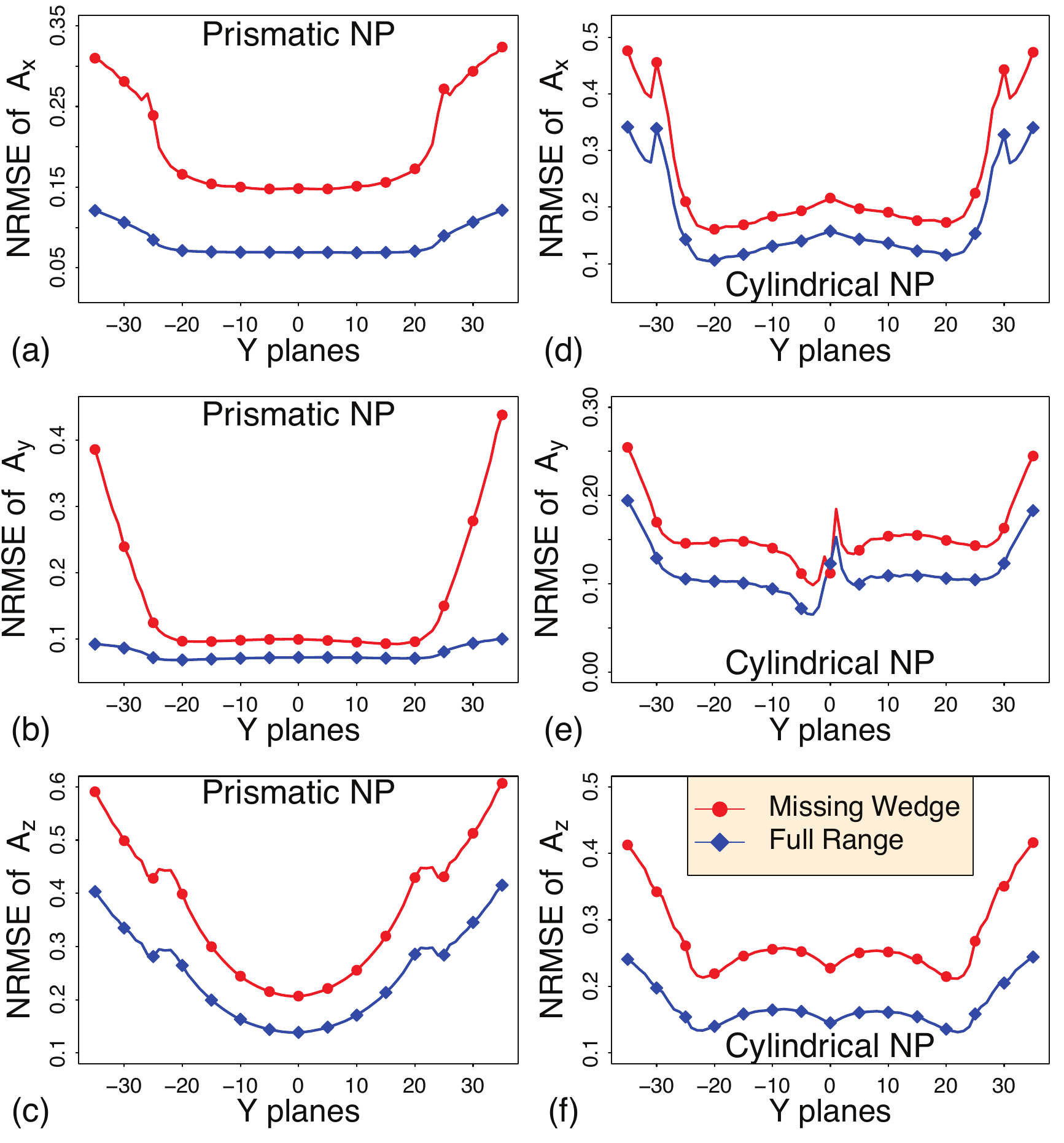}
\caption{Planar NRMSE plots of magnetic vector potential retrieved using the
VFET approach on the missing wedge projection set (line with circle)
and the full range projection set (line with diamond). The left column,
(a)-(c), shows NRMSE planar plots for the prismatic NP while
the right column, (d)-(e), shows the NRMSE planar plots for the cylindrical
NP. \label{fig:vfet_70vs90p_quantitative}}
\end{figure}

Fig.~\ref{fig:vfet_70vs90p_qualitative} compares sections of reconstructed vector potentials
with their ground truths, and lists the corresponding NRMSE values. 
Blurring at edges and ring artifacts are clearly more prominent in the reconstructions from the missing wedge data set, which has a higher NRMSE value than the reconstruction using the full tilt series.
Fig.~\ref{fig:vfet_70vs90p_quantitative} shows the NRMSE values corresponding to the $y$ planes in the range $[-35,35]$.
These plots indicate that the reconstruction deduced from the missing wedge data set exhibits a higher level of deviation from the ground truth than its full range counterpart by about $5$--$25$\%.  In the following section, we will replace the filter based analytical relation of eq.~\ref{eq:Ar_fbp_formula} with a Bayesian statistics-based numerical relation. 

%%%%%%%%%%%%%%%%%%%%%%%%%%%%%
\section{The MBIR framework\label{sec:gen-MBIR-framework} }
%%%%%%%%%%%%%%%%%%%%%%%%%%%%%
We introduce the MBIR approach by first formulating its framework for a generic tomography problem. 
Let $x\in \mathbb{R}^{N}$ be the discrete vector of an unknown image and $y\in\mathbb{R}^{M}$ be the
discrete vector of projection measurements of $x$ at a variety of
angles. Then the MBIR approach performs a reconstruction by maximizing
the joint probability distribution resulting from the likelihood function, $\mathcal{P}(y|x)$,
and the prior distribution, $\mathcal{P}(x)$. This methodology is more
commonly known as the maximum-a-posteriori (MAP) estimation technique \cite{C_bouman_MAP_paper}.
Mathematically, the MAP estimate is expressed as:
\begin{equation}
\hat{x}_{\text{MAP}}=\underset{x}{\text{argmin}}\left\{ -\log\mathcal{P}(y|x)-\log\mathcal{P}(x)\right\} .\label{eq:map_basic_eq}
\end{equation}
We use the Poisson probability mass function to model the likelihood term. We use a second order Taylor
series expansion of the logarithm of the Poisson distribution \cite{C_bouman_taylor_expansion}
to approximate the first term of eq.\,\ref{eq:map_basic_eq} as:
\begin{equation}
\log\mathcal{P}(y|x)\approx-\frac{1}{2}(y-Hx)^{T}W(y-Hx)+f(y),\label{eq:likelihood_taylor_expansion}
\end{equation}
where $H$ is the forward projection matrix, $W$ is a diagonal noise
weighting matrix and $f(y)$ is a term independent of the optimization
variable $x$. Additionally, $H$ is an orthogonal matrix such that
$H^{T}=H^{-1}$; hence, $H^{T}$ denotes the back projection operator. 

The prior term in eq.~\ref{eq:map_basic_eq} is modeled using a Markov Random Field (MRF) in:
\begin{equation}
\log\mathcal{P}(x)=-\underset{\{i,j\}\in\mathcal{C}}{\sum}b_{ij}\rho(x_{i}-x_{j}),\label{eq:log_prior}
\end{equation}
where $\mathcal{C}$ denotes the set of neighboring pixels, $b_{ij}$ is a non-causal symmetric
weighing filter that is normalized such that $\sum b_{ij}=1$, and $\rho(\cdot)$ is the potential function. 
The choice of the MRF as the prior model is due to its proven usefulness in image processing as well as in tomographic reconstructions \cite{MRF_denoise_app,MRF_tomo_app}. In the present paper, we resort to a class of MRFs called the $q$-Generalized
Gaussian Markov Random Field ($q$-GGMRF) to model the prior. Accordingly,
the potential function in eq.~\ref{eq:log_prior} is given by:
\begin{equation}
\rho(\Delta)=\frac{\left|\Delta\right|^{p}}{p\sigma_{x}^{p}}\left(\frac{\left|\frac{\Delta}{T\sigma_{x}}\right|^{q-p}}{1+\left|\frac{\Delta}{T\sigma_{x}}\right|^{q-p}}\right),\label{eq:GGMRF}
\end{equation}
where $\Delta=x_{i}-x_{j}$, and $p,\ q,\ \sigma_{x}\text{ and }T$ are the $q$-GGMRF parameters. Typically, $1\leq p\leq q\leq 2$ is used to ensure strict convexity of the potential function and, subsequently, of the MAP optimization \cite{Venk_mbir_haadf_stem}; in this paper we use $q=2$. When $p$ is set to $2$, the potential function is quadratic and the prior model facilitates a reconstruction with smooth edges. On the other hand, when $p$ is close to $1$, the prior model performs sharp edge preserving reconstructions. Similarly, $\sigma_{x}$ is the variance of the prior distribution and its value is set to achieve a balance between noise and resolution. Finally, the constant $T$ determines the approximate threshold of transition between low and high contrast regions.

Substituting the log-likelihood expression of eq.~\ref{eq:likelihood_taylor_expansion} and the prior model of eq.~\ref{eq:log_prior} into eq.~\ref{eq:map_basic_eq}, and considering $W$ as the identity matrix,  one obtains the MAP reconstruction to be the solution of the following optimization problem:
\begin{equation}
    \hat{x}_{\text{MAP}}=\underset{x}{\text{argmin}}\left\{ \frac{1}{2}\left\Vert y-Hx\right\Vert ^{2}+\underset{\{i,j\}\in\mathcal{C}}{\sum}b_{ij}\rho(x_{i}-x_{j})\right\} ,\label{eq:MAP_interim}
\end{equation}
where the cost (objective) function being minimized is given by
\begin{equation}
    c(x)=\frac{1}{2}\left\Vert y-Hx\right\Vert^{2}+
    \underset{\{i,j\}\in\mathcal{C}}{\sum}b_{ij}\rho(x_{i}-x_{j}).\label{eq:cost_function_first}
\end{equation}

\begin{figure}[t]
\centering\leavevmode
\includegraphics[width=0.5\textwidth]{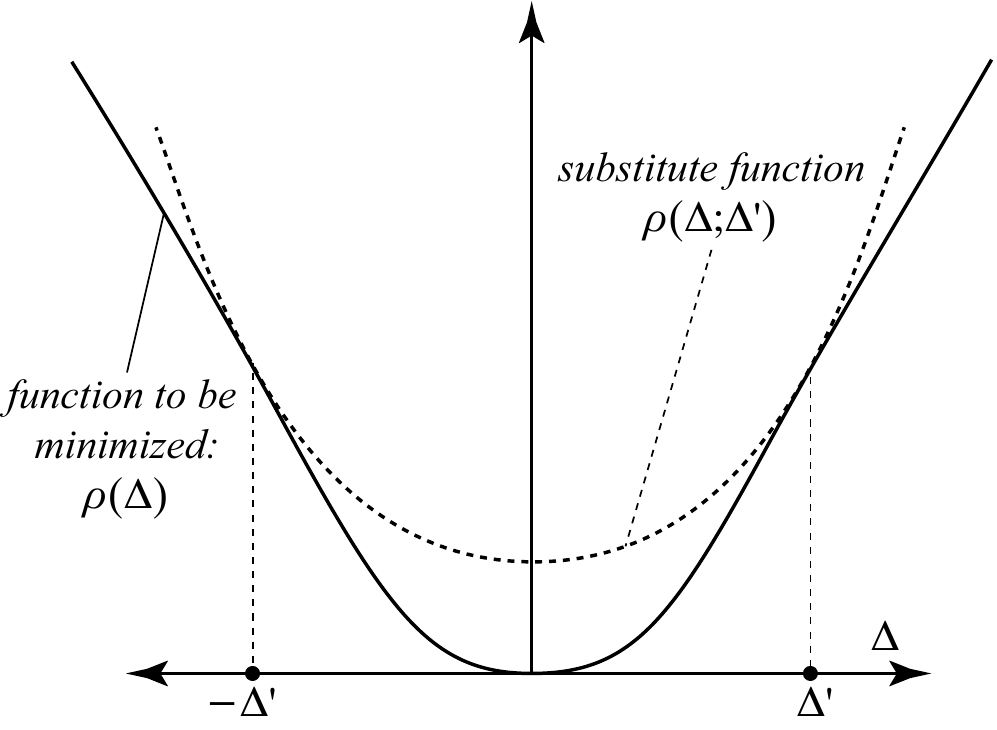}
\caption{Schematic of the intuition behind the formulation of the substitute
function of the non-quadratic potential function\label{fig:surrogate_fig}}
\end{figure}

Before proceeding with the minimization of the cost function, it should be noted that differentiating the cost function with respect to $x_{i}$ and setting the derivative to zero does not yield a closed form solution. Accordingly, one needs to make use of numerical methods, e.g., the Newton-Raphson method or the line search method, to determine the MAP estimate \cite{C_bouman_taylor_expansion,c_bouman_newton_m}. However, these numerical methods can be computationally expensive to implement. Thus, we introduce the substitute  or surrogate function to derive a closed form solution of the MAP estimate and subsequently facilitate a faster implementation \cite{fast_imp_using_substitute_func}. A visual depiction of the intuition behind the substitute function is shown in Fig.~\ref{fig:surrogate_fig}. In accordance to the figure, symmetric bond method is employed to construct a substitute function,
$\rho(\Delta;\Delta')$, as an upper bound to the original function,
$\rho(\Delta)$, such that minimizing the substitute function results
in a lower value of the original function \cite{aditya_timbir}. Accordingly,
the potential function in eq. \ref{eq:log_prior} is upper bounded
by a symmetric and quadratic function of $\Delta$. Then its substitute
form can be expressed as: 
\begin{equation}
    \rho(\Delta;\Delta')=\frac{a_{2}}{2}\Delta^{2},\label{eq:substitube_func_begin}
\end{equation}
where $a_{2}$ is a function of $\Delta'$. Here, $a_{2}$ is determined by matching the gradient of $\rho(\Delta)$ and $\rho(\Delta;\Delta')$ at $\Delta=\Delta'$ to yield:
\begin{equation}
    a_{2}=\frac{\rho'(\Delta')}{\Delta'}.\label{eq:a2_value}
\end{equation}
Substituting the value of $a_{2}$ in eq.~\ref{eq:substitube_func_begin}
results in the following symmetric bound substitute function:
\begin{equation}
    \rho(\Delta;\Delta')=\begin{cases}
    \begin{array}{c}
    \frac{\rho'(\Delta')}{2\Delta'}\Delta^{2}\\
    \frac{\rho''(0)}{2}\Delta^{2}
    \end{array} & \begin{array}{c}
    \text{if }\Delta\neq0\\
    \text{if }\Delta=0
    \end{array}\end{cases},\label{eq:generalized_substitute_func}
\end{equation}
where the limiting value of $a_{2}=\rho''(0)$ is used when $\Delta=0$. Subsequently, we obtain the new MAP cost function as:
\begin{equation}
    c(x;x')=\frac{1}{2}\left\Vert y-Hx\right\Vert^{2}+
    \underset{\{i,j\}\in\mathcal{C}}{\sum}\tilde{b}_{ij}(x_{i}-x_{j})^{2},\label{eq:surrogate_cost}
\end{equation}
where
\begin{alignat}{1}
    \tilde{b}_{ij} & =\begin{cases}
    b_{ij}\frac{|x'_{i}-x'_{j}|^{p-2}}{2\sigma_{x}^{p}}\frac{\left|\frac{x'_{i}-x'_{j}}{T\sigma_{x}}\right|^{q-p}\left(\frac{q}{p}+\left|\frac{x'_{i}-x'_{j}}{T\sigma_{x}}\right|^{q-p}\right)}{\left(1+\left|\frac{x'_{i}-x'_{j}}{T\sigma_{x}}\right|^{q-p}\right)^{2}} & \text{for }\Delta\ne0\\
    b_{ij}\frac{1}{p\sigma_{x}^{p}} & \text{for }\Delta=0
    \end{cases}.\label{eq:tilde_bij_val}
\end{alignat}
The surrogate cost function in eq.~\ref{eq:surrogate_cost} is quadratic and so its minimum can be expressed in closed form.  We use the Iterative Coordinate Descent (ICD) method to numerically minimize the surrogate cost function \cite{fast_imp_using_substitute_func}. ICD works by sequentially minimizing the cost function of eq.~\ref{eq:surrogate_cost} with respect to each pixel. The convergence of the ICD technique to solve the optimization problem can be inferred from the fact that the surrogate cost function in eq.~\ref{eq:surrogate_cost} is strictly convex so that any locally optimal point is also globally optimal.  Thus, we proceed to differentiate the surrogate cost function and
solve for the minimum $x_{i}$ to yield each pixel update as:
\begin{equation}
    x_{i}\leftarrow\frac{H^{T}y+2\underset{\{i,j\}\epsilon\mathcal{C}}{\sum}\tilde{b}_{ij}x_{j}}{1+2\underset{\{i,j\}\epsilon C}{\sum}\tilde{b}_{ij}}.\label{eq:xi_update}
\end{equation}

\begin{algorithm}[t]
\begin{framed}
\begin{algorithmic}[1]
\State initialize $u\leftarrow\text{FBP}(y)$
\State  $v\leftarrow H^{T}y$
\While{not converged}
\For{$i=1$ to $N$}
\State $\text{for }\{i,j\}\in\mathcal{C}\text{ of }u_{i}\text{ determine }\tilde{b}_{ij}\text{ using eq.~\ref{eq:tilde_bij_val}}$
\State $u_{i}\leftarrow (v_{i}+2\underset{\{i,j\}\in\mathcal{C}}{\sum}\tilde{b}_{ij}u_{j})/(1+2\underset{\{i,j\}\in C}{\sum}\tilde{b}_{ij})$
\EndFor
\State $e\leftarrow y-Hu$
\State $u\leftarrow u+H^{T}e$
\EndWhile
\end{algorithmic}
\caption{Fourier Transform-based MBIR method for scalar tomography\label{alg:general_mbir}}
\end{framed}
\end{algorithm}

%%%%%%%%%%%
% edited to here by MDG on 3/12/17
%%%%%%%%%%%

A pseudocode to minimize the surrogate cost function for a pixel wise update using the ICD technique is listed in Algorithm \ref{alg:general_mbir}. Note that, in contrast to the widely observed MBIR practice of updating the error sinogram, $e$, after each pixel update, we resort, instead, to updating $e$ after all $N$ pixels have been modified. This is primarily due to the fact that we will be using a Fourier-based forward model instead of a Radon based forward model \cite{qGGMRF_proof} to minimize the
cost function. The Fourier-based forward model is chosen to make the framework compatible with imposing a Coulomb gauge constraint when we extend it for the reconstruction of $\mathbf{A}(\mathbf{r})$. 

\begin{figure}[h]
\centering\leavevmode
\includegraphics[width=0.4\textwidth]{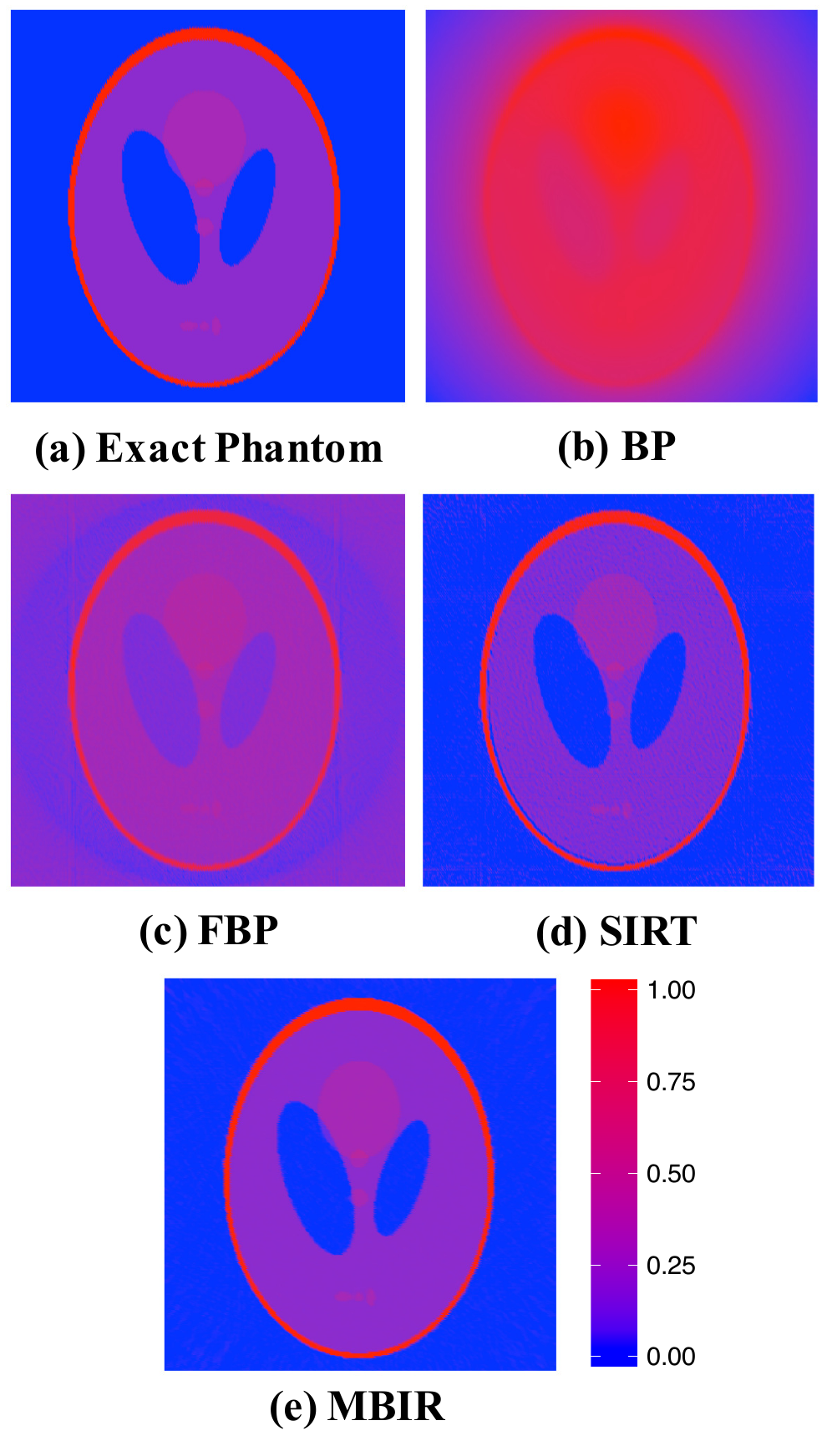}
\caption{Illustration of reconstructions from different tomographic methods
for the purpose of qualitative, (b)-(e), and quantitative, (f), comparisons.
The ground truth Shepp-Logan phantom (a) is projected in an angular
range of $\pm 90^{\circ}$ and reconstructed using several tomographic approaches:
(b) Back Projection (BP); (c) Filtered Back Projection (FBP);
(d) Simultaneous Iterative Reconstruction Technique (SIRT); and (e) Model
Based Iterative Reconstruction (MBIR). \label{fig:scalar_tomo_methods}}
\end{figure}

We conclude this section with the implementation of the MBIR methodology to a 2D scalar tomography problem. We make use of the standard Shepp-Logan head phantom \cite{Shepp-Logan} to illustrate our results. The phantom or ground truth is depicted in Fig.~\ref{fig:scalar_tomo_methods}(a); it is $[256\times256]$ pixels in size and has values ranging from $0$ to $1$, which we map on a blue-red color scale. In addition to the result from the MBIR approach (Fig.~\ref{fig:scalar_tomo_methods}(e)), we also show results from other reconstruction techniques, such as back projection (BP) in Fig.\,\ref{fig:scalar_tomo_methods}(b) and FBP in Fig.\,\ref{fig:scalar_tomo_methods}(c). We used a band-limited Ram-Lak ramp filter as our filtering function
\cite{ram-lak-paper,ramp_filter}. The result from an implementation of a conventional iterative tomography method known as the Simultaneous Iterative Reconstruction Technique (SIRT) is shown in Fig.\,\ref{fig:scalar_tomo_methods}(d). 

\begin{table}[htbp]
\caption{\label{tb:RMSE}Root Mean Square Error (RMSE) analysis between the Shepp-Logan phantom and the reconstructions for different tomographic approaches.}
\leavevmode\centering
\begin{tabular}{lc}
\hline
\textbf{Tomographic Method} & \textbf{RMSE}\\
\hline
Back-Projection (BP) & $0.4635$\\
Filtered Back-Projection (FBP) & $0.1536$\\
Simultaneous Iterative Reconstruction Technique (SIRT) & $0.0793$ \\
Model-Based Iterative Reconstruction (MBIR) & $0.0213$\\
\hline
\end{tabular}
\end{table}

Table~\ref{tb:RMSE} lists the root mean square error (RMSE) between the reconstructed results, $\phi$, and
the true phantom, $\hat{\phi}$, for several tomographic reconstruction methods. Mathematically, the RMSE is expressed as:
\begin{equation}
    \text{RMSE}=\sqrt{\frac{1}{N}\underset{i=1}{\overset{N}{\sum}}(\hat{\phi_{i}}-\phi_{i})^{2}}.\label{eq:RMSE}
\end{equation}
The RMSE value of the BP result was $0.4635$; such a high RMSE value was expected as the BP method is the simplest form of reconstruction technique and does not incorporate any filter or iterative technique in its model. The RMSE value decreases to about $0.1586$ for the FBP approach. Reconstruction by the SIRT method shows further improvement as the ring artifact, seen in the FBP result, is entirely eliminated, resulting in an RMSE value of $0.0793$. The effectiveness of the SIRT approach
over analytical method like the FBP approach stems from the fact that the SIRT algorithm iteratively minimize the error sinogram as \cite{SIRT_update}:
\begin{equation}
    x^{(\text{new})}\leftarrow x^{(\text{old})}+\lambda H^{T}W(y-Hx^{(\text{old})})\label{eq:SIRT_update}
\end{equation}
where $\lambda>0$ is a relaxation parameter. However, the ill-posed nature of the tomography problem limits the SIRT strategy to exhibit only semi-convergence \cite{SIRT_update,semi_convergence}. Accordingly, improvements seen in the first few SIRT iterations start to deteriorate as the number of iterations increases due to noise propagation. In our analysis, we show a locally converged SIRT result obtained after $10$ iterations with $\lambda=0.25$ and $W$ set as the identity matrix.

\begin{figure}[tph]
\centering\leavevmode
\includegraphics[width=0.5\textwidth]{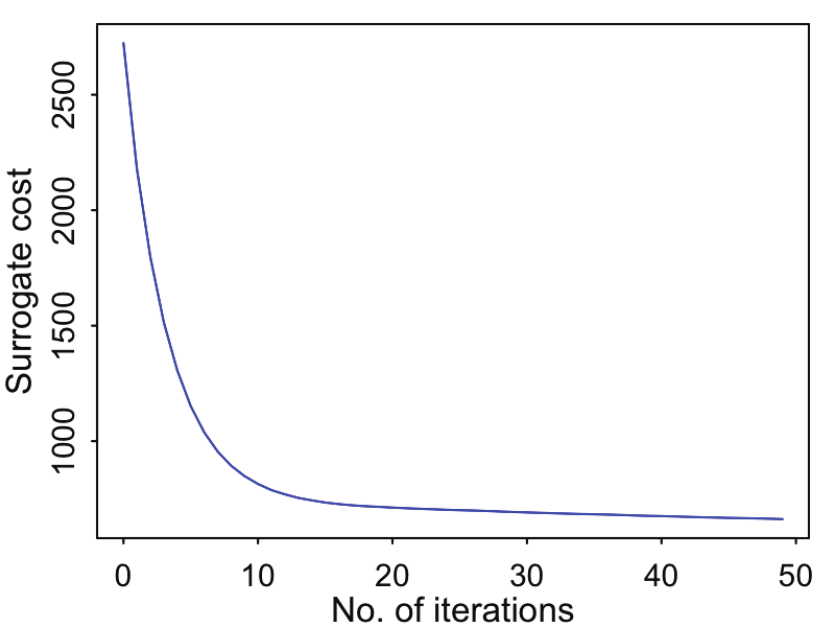}
\caption{Illustration of minimization of surrogate cost function using the ICD technique to deduce the MBIR estimate of reconstructed Shepp-Logan phantom in Fig.\,\ref{fig:scalar_tomo_methods}(e)\label{fig:mbir_2d_cost}}
\end{figure}

The MBIR reconstruction is obtained by iteratively minimizing the surrogate cost function of eq.\,\ref{eq:surrogate_cost} which, in turn, is achieved by following Algorithm \ref{alg:general_mbir}. The $q$-GGMRF parameters used in this implementation were $p=1.1,\ T=0.001\text{ and }\sigma_{x}=0.8$. We used a $3\times3$ non-causal weighting matrix, $b_{ij}$, to incorporate the influence of two nearest neighbors for any given pixel. The first nearest neighbors are weighted by a factor of $1/6$ while the second nearest nearest neighbors are weighted by a factor of $1/12$. The decreasing trend of the surrogate cost function over several iterations is shown in Fig.~\ref{fig:mbir_2d_cost}. An important point to note here is that the decreasing trend of the surrogate cost is a direct consequence of its strictly convex nature. Convergence after the cost minimization means that the reconstruction is globally minimal and unique. Hence, the coarseness seen along the inner region of the locally converged SIRT result is no longer visible in the MBIR result. Also, the superiority of the MBIR approach over any of the other tomography methods can be discerned from the RMSE analysis which reveals an error of only $0.0213$.

%%%%%%%%%%%%%%%%%%%%%%%%%%%%%
\section{MBIR-based vector field reconstruction\label{sec:MBIR_vec_imp}}
%%%%%%%%%%%%%%%%%%%%%%%%%%%%%
The MBIR framework for vector field reconstruction revolves around solving an optimization problem similar in form to the one expressed in eq.\,\ref{eq:map_basic_eq}.  Specifically, we seek to formulate a framework that will iteratively work to minimize the sum of the squared differences between the data and its estimated forward projection, in combination with a regularizing prior function, $\mathcal{P}(x)$. However, in contrast to scalar tomography, vector reconstruction entails measurements from more than one tilt series; each tilt series involves a contribution from different components of $\mathbf{A}(\mathbf{r})$. We also impose a gauge constraint to reconstruct all three components of vector potential from just two tilt series. Accordingly, our generic MBIR framework from section~\ref{sec:gen-MBIR-framework} needs to be expanded to account for input measurements from two tilt
series and imposition of a gauge constraint. 

We begin MBIR based vector reconstruction by representing the input data by $\varphi_{m}=\left[\varphi_{m,x},\varphi_{m,y}\right]$, and subsequent estimates of the magnetic vector potential by $\mathbf{x}=\left[x_{x},x_{y},x_{z}\right]$. Let $\mathbf{H}=\left[H_{x},H_{y}\right]$ where $H_{x}$ and $H_{y}$ are operators for the forward projection of the magnetic vector potential for the $x$ and $y$ tilt series (counter-clockwise), in accordance with eqs.~\ref{eq:x_tilt_fst} and \ref{eq:y_tilt_fst}, respectively. Then the forward projection of $\mathbf{x}$ yields, $\mathbf{Hx}=\left[H_{x},H_{y}\right]\mathbf{x}=\left[U_{x},U_{y}\right]$.
We define a deconvolution operator, $\mathbf{\mathcal{\mathbf{D}}}=[D_{x},D_{y},D_{z}]$, which will serve to evaluate the contributions of the magnetic phase shift (eqs.~\ref{eq:phi_x_contribution_Ar} and \ref{eq:phi_y_contribution_Ar}) to each component of vector potential $\mathbf{A}(\mathbf{r})$ separately. Thus any of the components $D_{*}$ of the deconvolution operator, when applied to the phase shift, yields $D_{*}\varphi_{m}=\left[D_{*}\varphi_{m,x},D_{*}\varphi_{m,y}\right]= \left[\varphi_{m,x,*},\varphi_{m,x,*}\right]$; and its application to $\mathbf{A}(\mathbf{r})$ yields $D_{*}\mathbf{A}(\mathbf{r})=A_{*}$.

Next, we define the prior model analogous to the one defined in section~\ref{sec:VFET_framework}. We use a $q$-GGMRF as the potential function that minimizes the cost function for a given voxel based on the difference with its neighboring voxels. However, contrary to the 2D MBIR approach, we incorporate the influence of three nearest neighbors for any given voxel while evaluating the potential function $\rho(\cdot)$. Accordingly, the non-causal weighting matrix, $b_{ij}$, now contains $3\times3\times3$ values with first, second and third weighing factors assigned as, $9/132$, $9/264$, and $9/396$, respectively. Finally, we again make use of the surrogate majorization technique such that the $q$-GGMRF potential function assumes a quadratic form for $p\neq2$. 

Having defined all the pertinent variables, we can now express MBIR-based vector reconstruction as the solution to the
following optimization problem:
\begin{equation}
    \hat{\mathbf{x}}=\underset{\mathbf{x}}{\text{argmin}}\left\{ \frac{1}{2}\left\Vert \mathbf{D}\varphi_{m}-\mathbf{D(Hx)}\right\Vert ^{2}+\underset{\{i,j\}\in\mathcal{C}}{\sum}\tilde{b}_{ij}\mathbf{D}(\mathbf{x}(i)-\mathbf{x}(j))^{2}\right\} \label{eq:generic_MAP_4_vec_recon}
\end{equation}
where $\tilde{b}_{ij}$ is determined using eq.\,\ref{eq:tilde_bij_val}. In practice, we do not directly solve eq.~\ref{eq:generic_MAP_4_vec_recon}. Instead, we make use of the deconvolution operator, $\mathbf{D}$, and de-convolve eq.~\ref{eq:generic_MAP_4_vec_recon} into three MAP estimation problems as follows:
\begin{align}
    \hat{x}_{x}&=\underset{\mathbf{x}}{\text{argmin}}\left\{ \frac{1}{2}\left\Vert D_{x}\varphi_{m}-D_{x}\mathbf{(Hx)}\right\Vert ^{2}+\underset{\{i,j\}\in\mathcal{C}}{\sum}\tilde{b}_{ij}D_{x}(\mathbf{x}(i)-\mathbf{x}(j))^{2}\right\} ,\label{eq:MAP_est_Ax}\\
    \hat{x}_{y}&=\underset{\mathbf{x}}{\text{argmin}}\left\{ \frac{1}{2}\left\Vert D_{y}\varphi_{m}-D_{y}\mathbf{(Hx)}\right\Vert ^{2}+\underset{\{i,j\}\in\mathcal{C}}{\sum}\tilde{b}_{ij}D_{y}(\mathbf{x}(i)-\mathbf{x}(j))^{2}\right\} ,\label{eq:MAP_est_Ay}\\
    \hat{x}_{z}&=\underset{\mathbf{x}}{\text{argmin}}\left\{ \frac{1}{2}\left\Vert D_{z}\varphi_{m}-D_{z}\mathbf{(Hx)}\right\Vert ^{2}+\underset{\{i,j\}\in\mathcal{C}}{\sum}\tilde{b}_{ij}D_{z}(\mathbf{x}(i)-\mathbf{x}(j))^{2}\right\} .\label{eq:MAP_est_Az}
\end{align}
Next, we identify the cost function associated with each of the potential estimates in eqs.~\ref{eq:MAP_est_Ax}, \ref{eq:MAP_est_Ay}, and \ref{eq:MAP_est_Az}. For instance, the surrogate cost associated with the MAP estimate of $x_{x}$ is expressed as:
\begin{equation}
    c(\mathbf{x},x_{x};x_{x}')=\frac{1}{2}\left\Vert D_{x}\varphi_{m}-D_{x}\mathbf{(Hx)}\right\Vert ^{2}+\underset{\{i,j\}\in\mathcal{C}}{\sum}\tilde{b}_{ij}D_{x}(\mathbf{x}(i)-\mathbf{x}(j))^{2}.
    \label{eq:surrogate_cost_Ax}
\end{equation}
Then we use the ICD algorithm, as detailed in section~\ref{sec:gen-MBIR-framework}, to minimize the surrogate cost function of eq.~\ref{eq:surrogate_cost_Ax}. Differentiating eq.~\ref{eq:surrogate_cost_Ax} with respect to $\mathbf{x}(i)$ and setting the result equal to zero, we obtain:
\begin{equation}
    x_{x}(i)\leftarrow\frac{\left(H_{x}^{T}\varphi_{m,x,x}+H_{y}^{T}\varphi_{m,y,x}\right)+2\underset{\{i,j\}\in\mathcal{C}}{\sum}\tilde{b}_{ij}x_{x}(j)}{1+2\underset{\{i,j\}\in\mathcal{C}}{\sum}\tilde{b}_{ij}}.\label{eq:icd_Ax}
\end{equation}
In a similar manner, we solve for minimum $\mathbf{x}(i)$ corresponding to surrogate the cost functions of $\hat{x}_{y}$ and $\hat{x}_{z}$ in eqs.\,\ref{eq:MAP_est_Ay} and \ref{eq:MAP_est_Az} to obtain:
\begin{align}
    x_{y}(i)&\leftarrow\frac{\left(H_{x}^{T}\varphi_{m,x,y}+H_{y}^{T}\varphi_{m,y,y}\right)+2\underset{\{i,j\}\in\mathcal{C}}{\sum}\tilde{b}_{ij}x_{y}(j)}{1+2\underset{\{i,j\}\in\mathcal{C}}{\sum}\tilde{b}_{ij}},\label{eq:ICD_Ay}\\
    x_{z}(i)&\leftarrow\frac{\left(H_{x}^{T}\varphi_{m,x,z}+H_{y}^{T}\varphi_{m,y,z}\right)+2\underset{\{i,j\}\in\mathcal{C}}{\sum}\tilde{b}_{ij}x_{z}(j)}{1+2\underset{\{i,j\}\in\mathcal{C}}{\sum}\tilde{b}_{ij}}.\label{eq:ICD_Az}
\end{align}

\begin{algorithm}[t]
\begin{framed}
\begin{algorithmic}[1]
\State Evaluate $\mathbf{A}(\mathbf{r})\leftarrow\text{VFET}$
\State $\mathbf{F}(\mathbf{r})\leftarrow\text{VFET}$
\While{not converged}
\For{$i=1$ to $N$}
    \State Determine corresponding $\tilde{b}_{ij}$ for $\{i,j\}\in\mathcal{C}$ of $A_{x}(i)$ using eq.~\ref{eq:tilde_bij_val}
    \State $A_{x}(i)\leftarrow\frac{F_{x}(i)+2\underset{\{i,j\}\in\mathcal{C}}{\sum}\tilde{b}_{ij}A_{x}(j)}{1+2\underset{\{i,j\}\in C}{\sum}\tilde{b}_{ij}}$
\EndFor
\State Update $A_{y}\text{ and }A_{z}$ in a similar manner similar
\State Forward project $\mathbf{A}(\mathbf{r})$ using eqs.~\ref{eq:x_tilt_fst} and \ref{eq:y_tilt_fst} to determine $U_{x}$ and $U_{y}$, respectively
\State Determine the error sinogram for the two series as:
\begin{align*}
    e\_\varphi_{m,x}&=\varphi_{m,x}-U_{x};\\
    e\_\varphi_{m,y}&=\varphi_{m,y}-U_{y}.
\end{align*}
\State Use the results of imposing gauge constraint from eqs.~\ref{eq:phi_x_contribution_Ar}
and \ref{eq:phi_y_contribution_Ar} on $e\_\varphi_{m,x}$ and $e\_\varphi_{m,y}$
to determine $e\_\varphi_{m,x,x},e\_\varphi_{m,x,y},e\_\varphi_{m,x,z},e\_\varphi_{m,y,x},e\_\varphi_{m,y,y}$
and $e\_\varphi_{m,y,z}$ 
\State Update the components of the vector potential as
\begin{align*}
A_{x}&\leftarrow A_{x}+H^{T}(e\_\varphi_{m,x,x}+e\_\varphi_{m,y,x}); \\
A_{y}&\leftarrow A_{y}+H^{T}(e\_\varphi_{m,x,y}+e\_\varphi_{m,y,y}); \\
A_{z}&\leftarrow A_{z}+H^{T}(e\_\varphi_{m,x,z}+e\_\varphi_{m,y,z}).
\end{align*}
\EndWhile
\end{algorithmic}
\caption{MBIR method to reconstruct magnetic vector potential\label{alg:MBIR_Ar_recon}}
\end{framed}
\end{algorithm}

Note that the MAP estimate of each component of the vector potential, $\hat{x}_{*}$, is determined by performing element-wise minimization of its associated cost function from eqs.~\ref{eq:MAP_est_Ax}-\ref{eq:MAP_est_Az} with respect to the overall magnetic vector potential, $\mathbf{x}(i)$, instead of the individual component, $x_{*}(i)$. This is primarily due to the fact that each component is reconstructed with the aid of two tilt series, $\varphi_{m,x}\text{ and }\varphi_{m,y}$, which in turn have contributions from all three components of the vector potential, $\mathbf{A}(\mathbf{r})$. Hence, each ICD iteration consists of simultaneously evaluating eq.~\ref{eq:icd_Ax}-\ref{eq:ICD_Az} and then making use of the forward model and the deconvolution operator to update the estimates of the vector component as:
\begin{equation}
    \mathbf{D}\mathbf{x}\leftarrow\mathbf{D}\mathbf{x}+\mathbf{H}^{T}(\mathbf{D}\varphi_{m}-\mathbf{D}(\mathbf{H}\mathbf{x})).\label{eq:ICD_update_vec}
\end{equation}
A summary of the MBIR-based $\mathbf{A}(\mathbf{r})$ reconstruction is illustrated in the form of pseudo code in Algorithm \ref{alg:MBIR_Ar_recon}. In the next two subsections we make use of this MBIR algorithm to reconstruct the magnetic vector potential of synthetic as well as experimental data sets.

\subsection{Synthetic Data Set}
\begin{figure}[t]
\centering\leavevmode
\includegraphics[width=0.5\textwidth]{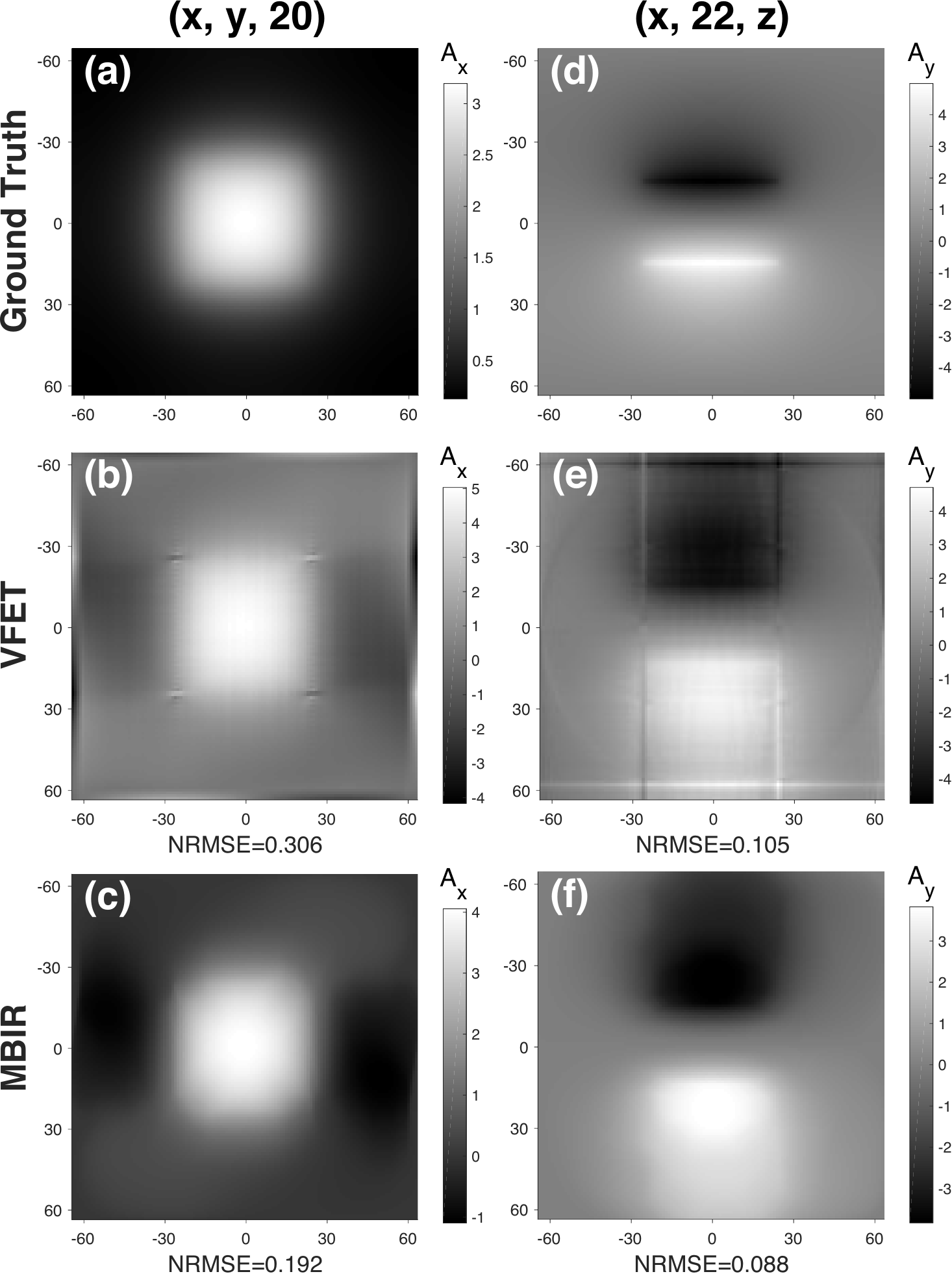}
\caption{Reconstructed magnetic vector potential of the prismatic NP deduced
from the VFET approach (center row) and the MBIR approach (bottom row)
with the aid of projections in the range $[-70^{\circ},70^{\circ}]$.
The top row depicts the ground truth. Plots in the left column
correspond to the plane $(x,y,20)$ while the ones in the right column
correspond to the plane $(x,22,z)$. \label{fig:cuboidal_MBIR_recon_qualitative}}
\end{figure}

\begin{figure}[t]
\centering\leavevmode
\includegraphics[width=0.5\textwidth]{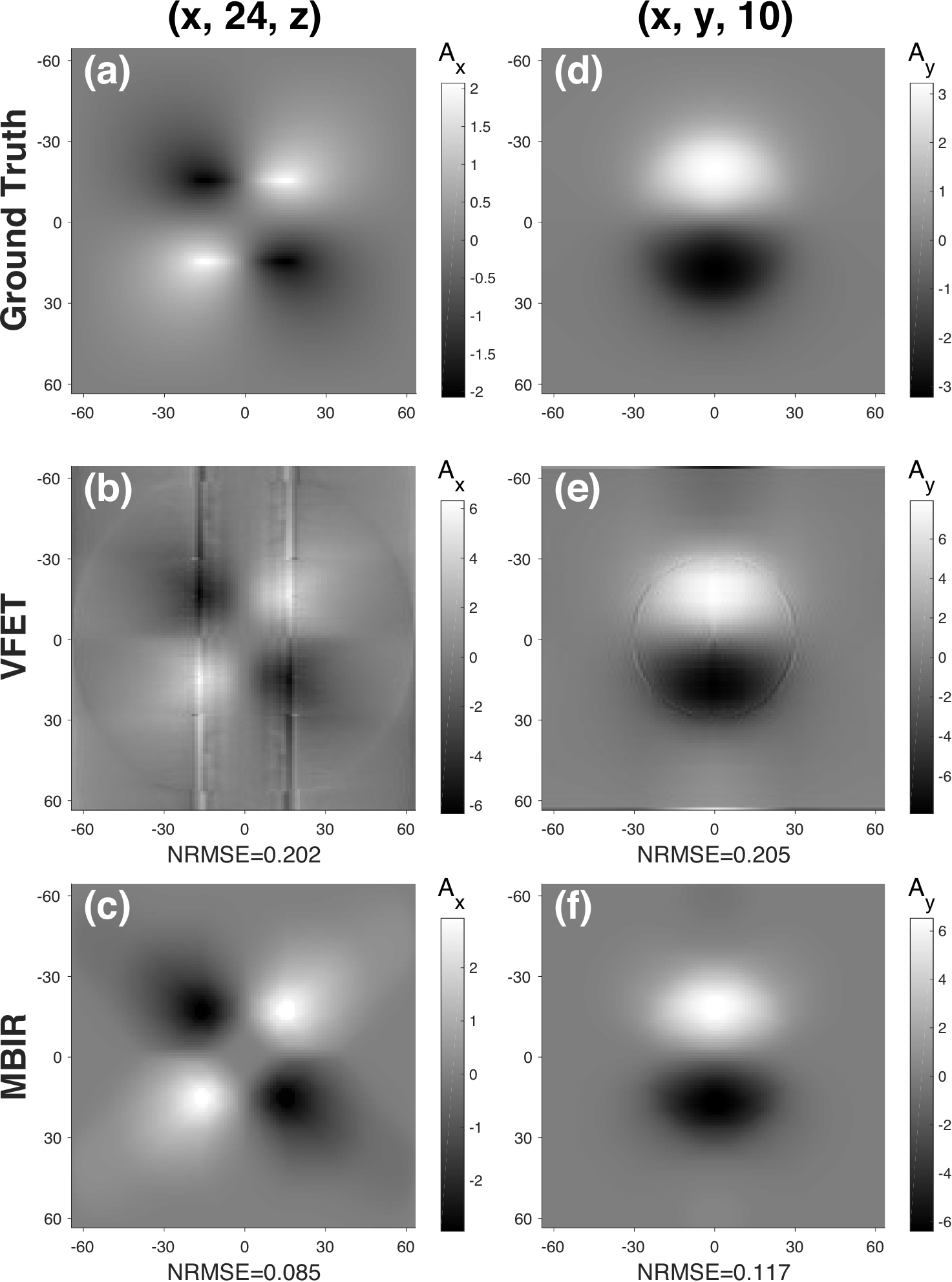}
\caption{Reconstructed magnetic vector potential of the cylindrical NP
from the VFET approach (center row) and the MBIR approach (bottom row)
with the aid of projection in the range $[-70^{\circ},70^{\circ}]$.
The top row depicts the ground truth. Plots in the left column
correspond to the plane $(x,24,z)$ while the ones in the right column
correspond to plane $(x,y,10)$. \label{fig:cylindrical_MBIR_qualitative}}
\end{figure}

\begin{figure}[t]
\centering\leavevmode
\includegraphics[width=0.5\textwidth]{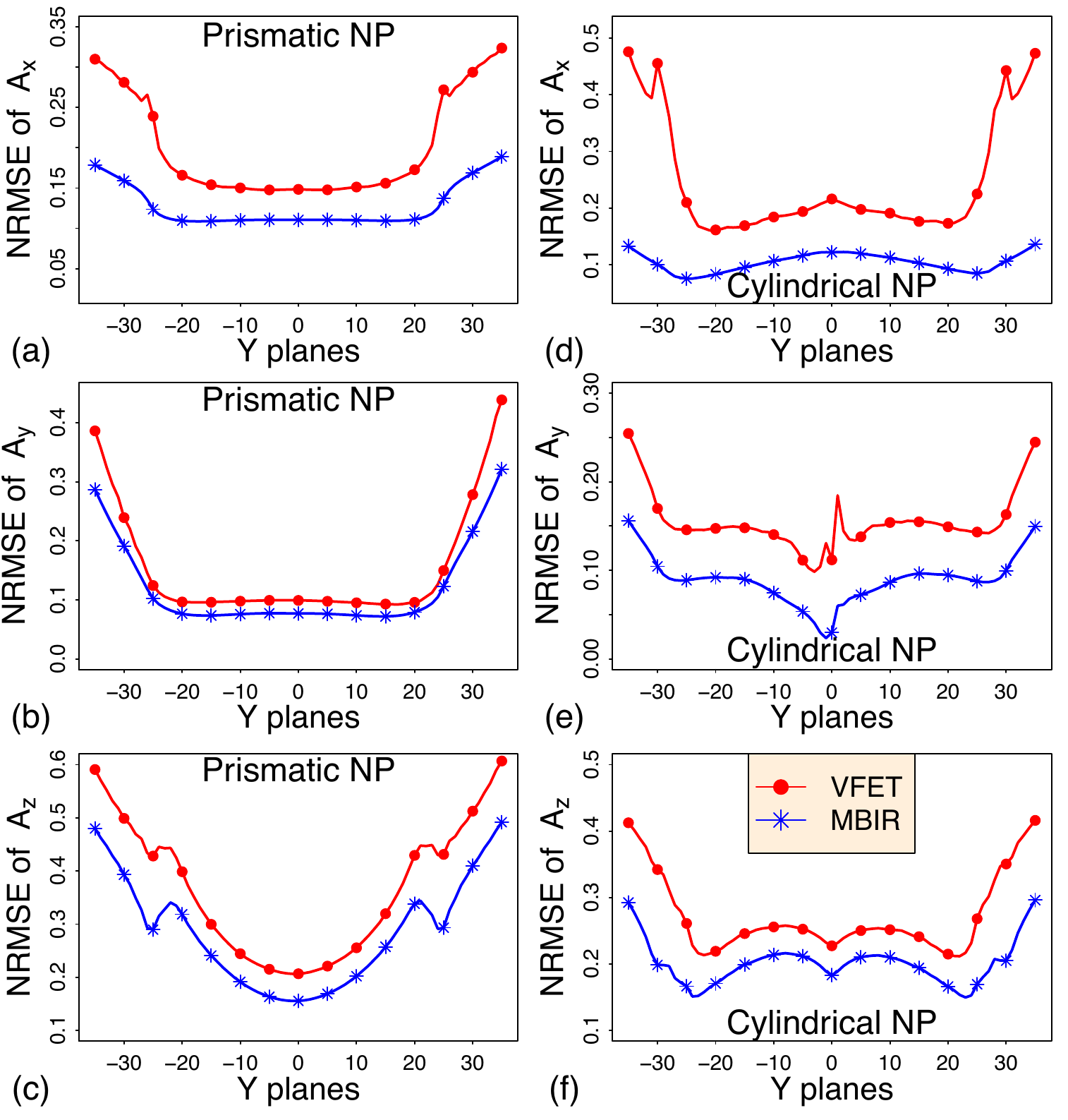}
\caption{Planar NRMSE plots of the magnetic vector potential retrieved using the
VFET approach (line with circle) and the MBIR approach (line with
asterisk). The left column shows the NRMSE planar plots
for the prismatic NP while the right column displays the NRMSE
planar plots for the cylindrical NP. \label{fig:mbir_Ar_quantitative}}
\end{figure}

In this section, we apply the MBIR algorithm to the reconstruction of $\mathbf{A}(\mathbf{r})$ for the cylindrical and prismatic MNPs discussed in section~\ref{sec:VFET_framework}. The reconstruction was performed making use of the missing wedge projection set since this set replicates the limited angular scenario typical of TEM based experimental measurements. The $q$-GGMRF parameters used for the reconstruction comprised of $p=1.001,\ T=0.01\text{ and }\sigma_{x}=0.8$. Selected results retrieved from the MBIR approach are depicted in Figs.~\ref{fig:cuboidal_MBIR_recon_qualitative} and \ref{fig:cylindrical_MBIR_qualitative}. These figures also include corresponding plots from the ground truth and the VFET approach for the purpose of qualitative comparison of reconstruction accuracy between the two approaches. We also present a quantitative comparison between the two methods in terms of NRMSE plots in Fig.\,\ref{fig:mbir_Ar_quantitative}.

A review of the plots in Figs.\,\ref{fig:vfet_70vs90p_qualitative}, \ref{fig:cuboidal_MBIR_recon_qualitative}, and \ref{fig:cylindrical_MBIR_qualitative} reveals that the low spatial resolutions, protrusions, ring artifacts and edge artifacts evident in the VFET reconstructions are significantly suppressed in the MBIR-based reconstructions. In case of the prismatic MNP, we note that the MBIR-based reconstruction displays values in closer proximity to the ground truth than the ones obtained from the VFET reconstructions, as shown in Figs.~\ref{fig:cuboidal_MBIR_recon_qualitative}(a-c). This is confirmed by the NRMSE plots in Figs.~\ref{fig:mbir_Ar_quantitative}a and \ref{fig:mbir_Ar_quantitative}c which show that the MBIR approach determines $A_{x}$ and $A_{z}$ with $10$ to $20\%$ less planar error (in $y$ axis) than those derived from the VFET approach. On planes, such as $(x,18,z)$, shown in Figs.~\ref{fig:vfet_70vs90p_qualitative}(a-c), where the reconstructed vector potential suffers from strong edge artifact as a direct consequence of the limited angular range of projection measurement, the MBIR reconstruction performs significantly better than the VFET approach. In Fig.~\ref{fig:cuboidal_MBIR_recon_qualitative}f, we observe that, although MBIR suppresses the ring artifact, the blurred edges persist; the loss of information due to the missing wedge is such that even an advanced algorithm as MBIR is unable to completely remove the artifacts observed in the VFET based reconstruction. This conclusion is also substantiated by the NRMSE plot in Fig.~\ref{fig:mbir_Ar_quantitative}b which reveals that the gain in reconstruction accuracy having adopted the MBIR approach is merely $1$ to $2\%$ compared to the VFET approach. 

For the cylindrical MNP, we see that the MBIR approach preserves the spatial resolution and retrieves vector potential values in closer proximity to the ground truth than the VFET approach, as depicted in Figs.~\ref{fig:cylindrical_MBIR_qualitative}(d-f). Accordingly, the NRMSE for $y$ planes of all three components of $\mathbf{A}(\mathbf{r})$ are less by about $10\%$ for MBIR reconstructions as compared to VFET reconstructions, as illustrated in Figs.~\ref{fig:mbir_Ar_quantitative}(d-f). Also, edge and ring artifacts observed in the VFET reconstruction are substantially diminished in the MBIR reconstruction, as shown in Fig.~\ref{fig:cylindrical_MBIR_qualitative}(a-c). However, both MBIR and VFET results exhibit protrusion effects due to the missing wedge. 

\subsection{Experimental Data Set}
In this final section we make use of a real data set to demonstrate the gain in reconstruction quality from making use of the MBIR approach. A 2D lattice of elongated $\text{Ni}_{80}\text{Fe}_{20}$ (permalloy/py) islands (``stadia'') was used for the comparative study of the MBIR and VFET techniques. The sample was fabricated on a JEOL $9300$ electron beam lithography system. A single layer of ZEP resist of $100$ nm thickness was coated on a Si/SiN substrate, followed by patterning of a square lattice with element shape parameters of $2\text{Lx}=290$ nm, $2\text{Ly}=130$ nm, and a lattice spacing of $a=390$ nm. A Py film of $20$ nm thickness was deposited on a seed layer of Cr ($3$ nm) using dc magnetron sputtering at $3$ mTorr pressure and $50$ W power. The pattern was transferred by a lift-off process. This was followed by optical lithography and wet-etching of Si to create electron transparent windows on $3$ mm square grids, which could be loaded directly into the TEM for observation. The microscopy was performed using the JEOL 2100F TEM equipped with a dedicated Lorentz lens and a spherical aberration corrector \cite{Py_sample_paper}. 

\begin{figure}[t]
\centering\leavevmode
\includegraphics[width=0.5\textwidth]{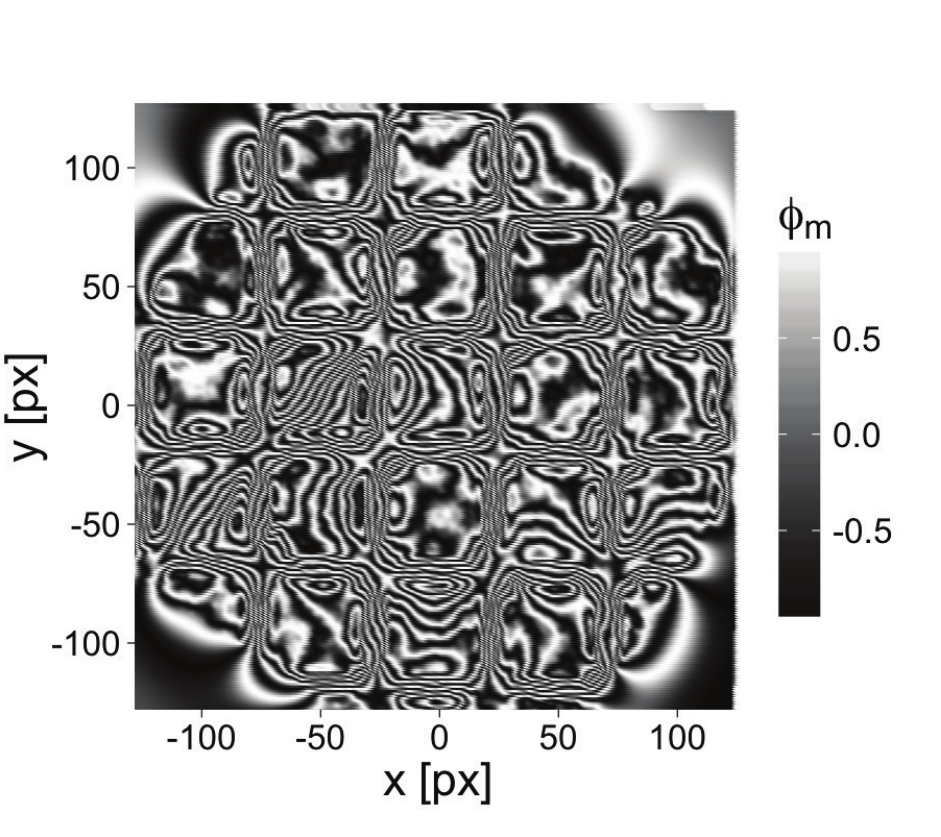}
\caption{Holographic contour map of the magnetic phase shift, $\cos(100\varphi_{m})$, of a Permalloy square lattice sample at $0^{\circ}$ tilt; the phase (in radians) has been multiplied by $100$ to enhance the contours and one pixel equals $6$ nm. \label{fig:Holographic-contour} }
\end{figure}

\begin{figure}[t]
\centering\leavevmode
\includegraphics[width=0.5\textwidth]{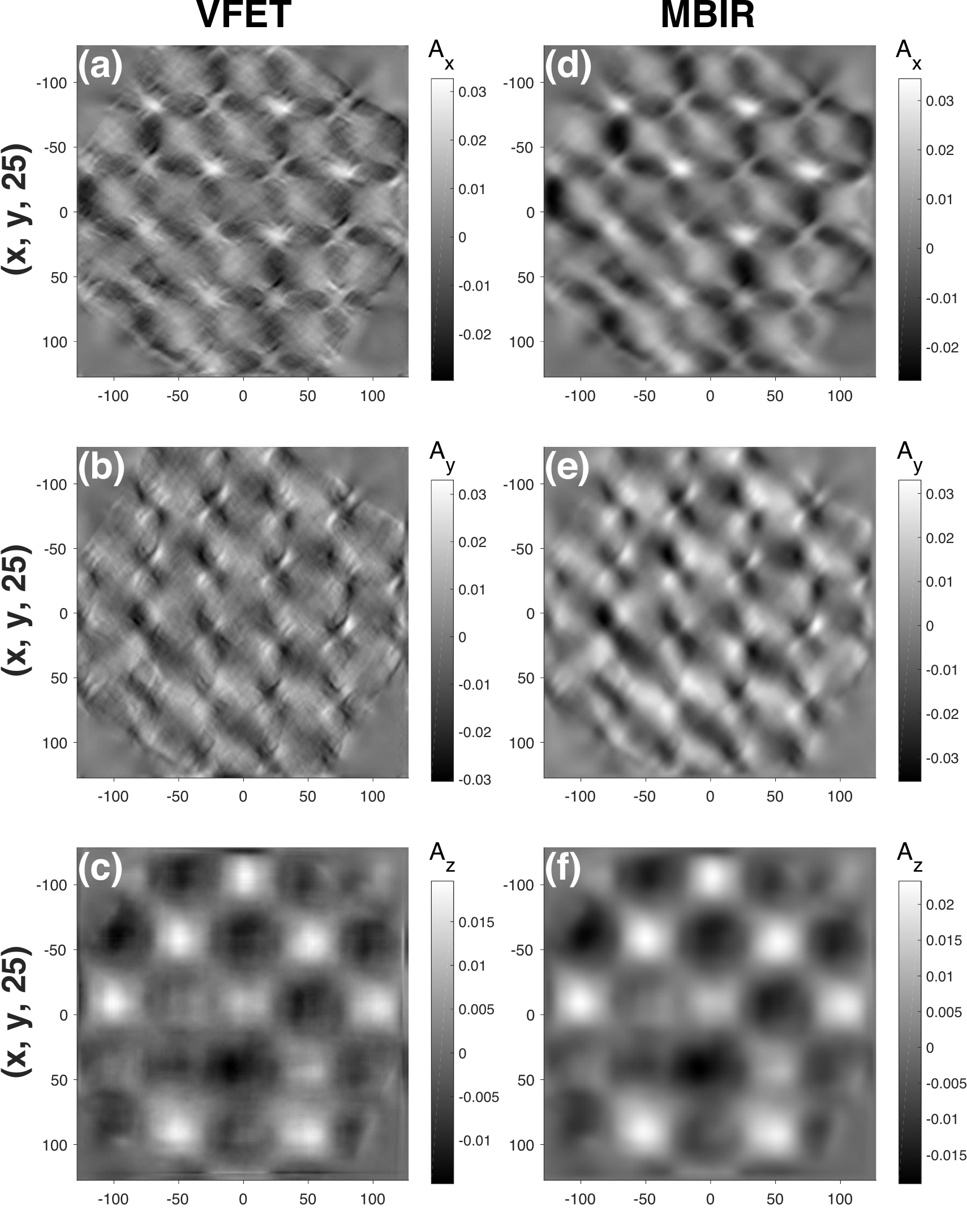}
\caption{Reconstructed magnetic vector potential, $\mathbf{A}(\mathbf{r})$,
of a Py square lattice from the VFET approach (left column) and the
MBIR approach (right column). The units for $\mathbf{A}(\mathbf{r})$
are T-px where $1$ pixel equals $6$ nm. \label{fig:py_vfet_mbir}}
\end{figure}

The projection measurements acquired for tomographic reconstruction consist of $x$ and $y$ tilt series with angles ranging from $-50^{\circ}$ to $+50^{\circ}$ at a step size of $1^{\circ}$. Fig.~\ref{fig:Holographic-contour} depicts one of the projections as a holographic plot, $\cos(100\varphi_m)$, of the magnetic phase shift of the sample at $0^{\circ}$ tilt. Each phase map has a resolution of $256\times256$ pixels ($6$ nm per pixel). We implemented our tomographic reconstruction on a 3D voxel grid with $256^{3}$ nodes. First, we employed the VFET approach (as discussed in section~\ref{sec:VFET_framework}) to determine an estimate of $\mathbf{A}(\mathbf{r})$ of the Py sample. These results are then used to initialize the MBIR algorithm. The $q$-GGMFR parameters used in the MBIR algorithm are $q=2.0$, $p=1.001$, $T=0.01$, and $\sigma_{x}=0.8$. The cost function in eq.~\ref{eq:generic_MAP_4_vec_recon} was then monotonically decreased over $35$ iterations to determine a MAP estimate of $\mathbf{A}(\mathbf{r})$ of the Py sample. The forward model calculation for each iteration was distributed over $24$ parallel threads using OpenMP and required about $18$ minutes to complete the iteration. Some of the $\mathbf{A}(\mathbf{r})$ results obtained from the VFET and  MBIR approaches are depicted in Fig.~\ref{fig:py_vfet_mbir}. 

A comparison of the plots deduced from the two methods reveals that the MBIR results show a significant gain in spatial resolution. The coarseness of the VFET result is substantially reduced in the MBIR result. Moreover, we see appropriately segmented and smoothly transitioning $\mathbf{A}(\mathbf{r})$ values in the MBIR results. The blurring artifacts observed in some of the VFET reconstruction sections are considerably diminished in the MBIR results. These reconstruction gains in the MBIR-based $\mathbf{A}(\mathbf{r})$ results are observed throughout the 3D spatial region when compared to their VFET counterparts. 

\begin{figure}[t]
\centering\leavevmode
\includegraphics[width=\textwidth]{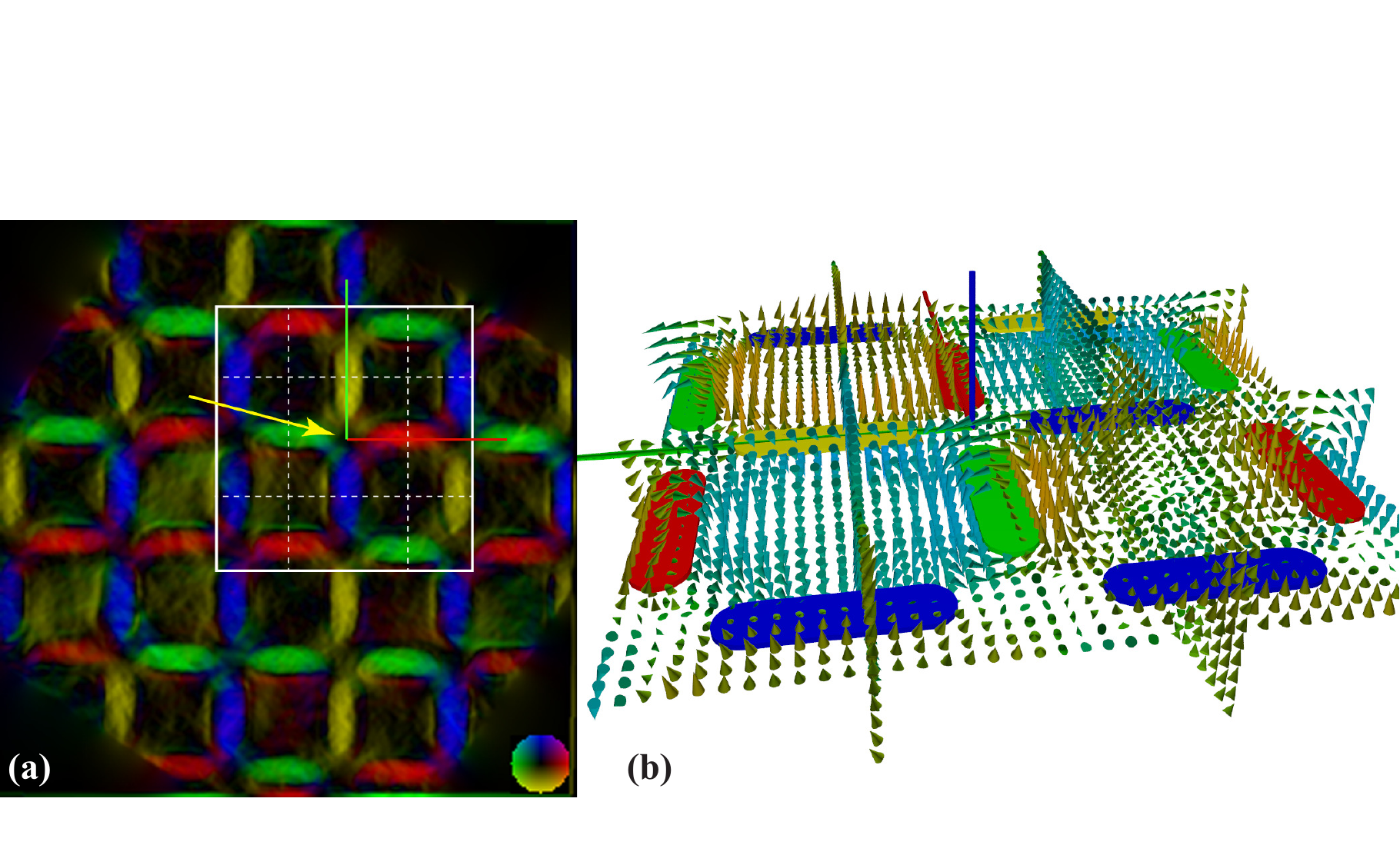}
\caption{(a) Magnetic induction map of the $0^{\circ}$ phase shift shown in
Fig.~\protect\ref{fig:Holographic-contour}. (b) Schematic of islands enclosed
inside the solid white lines and 3D magnetic vector potential corresponding
to the dotted lines in (a).\label{fig:3d_rendition}}
\end{figure}

Since the reconstructed quantity is a 3D vector field, we conclude this section with a 3D rendering (Fig.~\ref{fig:3d_rendition}) of the magnetic vector potential $\mathbf{A}(\mathbf{r})$ resulting from the MBIR reconstruction approach. Fig.~\ref{fig:3d_rendition}(a) shows a color-coded integrated (along the beam direction) magnetic induction map, derived from the magnetic phase shift by a gradient operation.  The white rectangle delineates the region used for the 3D rendering of the vector field in Fig.~\ref{fig:3d_rendition}(b). To reduce the complexity of the 3D rendering, the field vectors (shown as small cones with the appropriate orientation) are only drawn for the horizontal center plane going through the Py islands and four vertical planes corresponding to the dashed lines in Fig.~\ref{fig:3d_rendition}(a).  The yellow arrow indicates the viewing direction for the 3D rendering in Fig.~\ref{fig:3d_rendition}(b).  The individual islands are colored according to the in-plane direction of the integrated magnetic induction following the color wheel in (a).  Note that the induction directions, which are oriented along the local curl of the magnetic vector potential, are properly oriented; i.e., the opposite rotation of the vector field around the green and red islands is clearly visible, and application of the right-hand rule for the curl results in the correct direction of the integrated magnetic induction.  Furthermore, the square of islands on the left front of the image, and the two squares in the back, are all in a vortex state, i.e., the magnetization of the islands circulates clockwise or counterclockwise.  The corresponding section through the vector field shows vectors which are all blue ($\mathbf{A}$ pointing down) or all orange ($\mathbf{A}$ pointing up).  For the square on the front right, however, two of the islands are blue, so that there is no closed vortex and the vector field is much more complex in this area, reflecting the local frustrated state of the magnetization pattern.

\section{Conclusion}
In this contribution, we have incorporated the Bayesian inference-based MBIR technique into the tomographic reconstruction process for the 3D magnetic vector potential. The MBIR model is formulated by combining a vector slice-based forward model and a $q$-GGMRF-based prior model. An objective function was constructed and minimized using the Iterative Coordinate Descent technique to determine a MAP estimate of $\mathbf{A}(\mathbf{r})$. The MAP estimate shows diminished edge, ring and blurring artifacts when directly compared to reconstructions carried out with the more traditional filtered back-projection reconstruction from the VFET methodology. This qualitative conclusion is confirmed by the NRMSE values for simulated magnetic nan-particles with uniform
as well as non-uniform magnetization configurations and prismatic and cylindrical particle shapes. We have implemented the MBIR technique to reconstruct the magnetic vector potential $\mathbf{A}(\mathbf{r})$ of a 2D array of Py islands. The resulting vector field was compared to an integrated magnetic induction map derived from the magnetic phase shift and is in good agreement.  

\section*{Acknowledgements}
PK and MDG acknowledge financial support from a U.S. DOE grant (DE-FG02-01ER45893) which supported the development of the 2D MBIR approach, and an NSF grant (DMR \#1306296), which supported the final development of the 3D MBIR reconstructions. AM and CB acknowledge an AFOSR MURI grant \# FA9550-12-1-0458 for financial support. The experimental work by CP was supported by the U.S. Department of Energy, Office of Science, Basic Energy Sciences, Materials Sciences and Engineering Division. Use of Center for Nanoscale Materials was supported by the U.S. Department of Energy, Office of Science, Office of Basic Energy Sciences, under contract no. DE-AC02-06CH11357. The authors would also like to acknowledge the computational facilities of the Materials Characterization Facility at CMU under grant \# MCF-677785.

\newpage
\bibliographystyle{elsarticle-num}
\bibliography{MBIR-A}

\begin{thebibliography}{10}
\expandafter\ifx\csname url\endcsname\relax
  \def\url#1{\texttt{#1}}\fi
\expandafter\ifx\csname urlprefix\endcsname\relax\def\urlprefix{URL }\fi
\expandafter\ifx\csname href\endcsname\relax
  \def\href#1#2{#2} \def\path#1{#1}\fi

\bibitem{lorentz_force_jackson}
J.~Jackson, Classical Electrodynamics, second ed., Wiley, NY, 1975.

\bibitem{marc_editor_book}
M.~De~Graef, Lorentz Microscopy: Theoretical Basis and Image Simulations, M. D.
  Graef, and Y. Zhu (Eds.), Magnetic Imaging and Its Applications to Materials,
  Vol.~36, Academic Press, San Diego, CA, 2001 (Chapter 2).

\bibitem{A_B_paper}
Y.~Aharonov, D.~Bohm, Significance of electromagnetic potentials in the quantum
  theory, The Phys. Rev. 115 (1959) 485--491.

\bibitem{tomography_holography_review}
P.~A. Midgley, R.~E. Dunin-Borkowski, Electron tomography and holography in
  materials science, Nature Materials 8 (1997) 271--280.

\bibitem{TIE_paper}
D.~Paganin, K.~A. Nugent, Noninterferometric phase imaging with partially
  coherent light, Phys. Rev. Lett. 80 (1998) 2586--2589.

\bibitem{emma_TIE_paper}
E.~Humphrey, C.~Phatak, A.~Petford-Long, M.~De~Graef, Separation of
  electrostatic and magnetic phase shits using a modified
  transport-of-intensity equation, Ultramicroscopy 139 (2014) 5--12.

\bibitem{vfet_noise}
Z.~Kemp, T.~Petersen, D.~Paganin, K.~Spiers, M.~Weyland, M.~Morgan, Analysis of
  noise-induced errors in vector-field electron tomography, Phys. Rev. A 90
  (2014) 023859.

\bibitem{vfet_doppler}
S.~Lade, D.~Paganin, M.~Morgan, 3-{D} {V}ector tomography of
  doppler-transformed fields by filtered-backprojection, Opt. Commun. (2005)
  382--391.

\bibitem{vfet_three_tilt}
P.~Y. Rotha, M.~J. Morgan, D.~M. Paganin, Lorentz-electron vector tomography
  using two and three orthogonal tilt series, Phys. Rev. A 83 (2011) 023813.

\bibitem{e_tomo_review}
P.~A. Midgley, R.~E. Dunin-Borkowski, Electron tomography and holography in
  materials science, Nature Materials 8 (2009) 271--280.

\bibitem{c_bouman_local_strategy}
K.~Sauer, C.~Bouman, A {L}ocal {U}pdate {S}trategy for {I}terative
  {R}econstruction from {P}rojections, IEEE Trans. on Sig. Proc. 41 (1993)
  534--548.

\bibitem{qGGMRF_proof}
J.-B. Thibault, K.~D. Sauer, C.~A. Bouman, J.~Hsieh, A three-dimensional
  statistical approach to improved image quality for multislice helical {CT},
  Med. Phys. 34 (2007) 4526--4544.

\bibitem{norton_paper}
S.~J. Norton, Tomographic reconstruction of 2-{D} vector fields: application to
  flow imaging, Geophysical Journal 97 (1988) 161--168.

\bibitem{Juhin_paper}
P.~Juhin, Principles of doppler tomography, tech. report, center for
  mathematical sciences, lund institute of technology, SE-221 00 Lund, 1992.

\bibitem{vfet_cd}
C.~Phatak, M.~Beleggia, M.~De~Graef, Vector field electron tomography of
  magnetic materials: Theoretical development, Ultramicroscopy 108 (2008)
  503--513.

\bibitem{kak_slaney}
A.~C. Kak, M.~Slaney, Principles of Computerized Tomographic Imaging, Vol.~33,
  Cambridge University Press, Cambridge, UK, 2003.

\bibitem{philMagI_NP}
M.~Beleggia, Y.~Zhu, Electron optical phase shift of magnetic nanoparticles.
  {P}art {I}. basic concepts, Philosophical Magazine 83 (2003) 1045--1057.

\bibitem{emma_spherical_proj}
E.~Humphrey, M.~De~Graef, On the computation of the magnetic phase shift for
  magnetic nano-particles of arbitrary shape using a spherical projection
  model, Ultramicroscopy 129 (2013) 36--41.

\bibitem{C_bouman_MAP_paper}
C.~Bouman, K.~Sauer, A {G}eneralized {G}aussian {I}mage {M}odel for {E}dge
  {P}reserving {MAP} {E}stimation, IEEE Trans. on Img. Proc. 2 (1993) 296--310.

\bibitem{C_bouman_taylor_expansion}
C.~Bouman, K.~Sauer, A unified approach to statistical tomography using
  coordinate descent optimization, IEEE Trans. Img. Proc. 5 (1996) 480--492.

\bibitem{MRF_denoise_app}
H.~Derin, H.~Elliott, Modeling {S}egmentation of {N}oisy and {T}extured
  {I}mages {U}sing {G}ibbs {R}andom {F}ields, IEEE Transaction on Pattern
  Analysis and Machine Intelligence 9 (1987) 39--55.

\bibitem{MRF_tomo_app}
K.~Sauer, C.~Bouman, Baysian {E}stimation of {T}ransmission {T}omograms {U}sing
  {S}egmentation {B}ased {O}ptimization, IEEE Trans. on Nuclear Science 39
  (1992) 1144--1152.

\bibitem{Venk_mbir_haadf_stem}
S.~V. Venkatakrishnan, L.~F. Drummy, M.~A. Jackson, M.~De~Graef, J.~Simmons,
  C.~A. Bouman, A {M}odel {B}ased {I}terative {R}econstruction {A}lgorithm
  {F}or {H}igh {A}ngle {A}nnular {D}ark {F}ield - {S}canning {T}ransmission
  {E}lectron {M}icroscope ({HAADF-STEM}) {T}omography, IEEE Trans. on Img.
  Proc. 22 (2013) 4532--4543.

\bibitem{c_bouman_newton_m}
J.-B. Thibault, K.~Sauer, C.~Bouman, Newton-style optimization for emission
  tomographic estimation, J. of Elec. Imag. 9 (2000) 269--282.

\bibitem{fast_imp_using_substitute_func}
Z.~Yu, J.-B. Thibault, C.~A. Bouman, K.~D. Sauer, J.~Hsieh, Fast
  {M}odel-{B}ased {X-R}ay {CT R}econstruction {U}sing {S}patially
  {N}onhomogeneous {ICD} {O}ptimization, IEEE Trans. on Img. Proc. 20 (2011)
  161--175.

\bibitem{aditya_timbir}
K.~A. Mohan, S.~Venkatakrishnan, J.~W. Gibbs, E.~B. Gulsoy, X.~Xiao,
  M.~De~Graef, P.~W. Voorhees, C.~A. Bouman, {TIMBIR}: {A} {M}ethod for
  {T}ime-{S}pace {R}econstruction from {I}nterlaced {V}iews, IEEE Trans. on
  Comp. Img. 1 (2015) 96--111.

\bibitem{Shepp-Logan}
L.~A. Shepp, B.~F. Logan, The {F}ourier {R}econstruction of a {H}ead {S}ection,
  IEEE Trans. on Nuclear Science 21 (1974) 21--43.

\bibitem{ram-lak-paper}
G.~Ramachandran, A.~Lakshminarayanan, Three-dimensional reconstructions from
  radiographs and electron micrographs: {A}pplication of convolution instead of
  {F}ourier transforms, Proc Nat Acad Sci. 68 (1971) 2236--2240.

\bibitem{ramp_filter}
G.~L. Zeng, Revisit of the {R}amp {F}ilter, IEEE Trans. On Nuclear Science 62
  (2015) 131--136.

\bibitem{SIRT_update}
T.~Elfving, P.~C. Hansen, T.~Nikazad, Semi-convergence properties of
  {K}aczmarz's method, Inverse Problems 30 (2014) 055007.

\bibitem{semi_convergence}
T.~Elfving, P.~C. Hansen, T.~Nikazad, Semi-convergence and relaxation
  parameters for projected {SIRT A}lgorithms, SIAM J. Sci. Comput. 34 (2012)
  A2000?17.

\bibitem{Py_sample_paper}
C.~Phatak, A.~Petford-Long, O.~Heinonen, M.~Tanase, M.~De~Graef, Nanoscale
  structure of the magnetic induction at monopole defects in artificial
  spin-ice lattices, Phys. Rev. B 83 (2011) 174431.

\end{thebibliography}

\end{document}